\documentclass[12pt,a4paper]{article}
\usepackage[OT2,T1]{fontenc}
\usepackage{amssymb}
\usepackage{amsmath}
\usepackage{stmaryrd}
\usepackage{mathrsfs}
\usepackage{mathtools}
\usepackage{bm}
\usepackage[dvipdfmx]{graphicx}
\usepackage{bmpsize}
\usepackage{here}
\usepackage{subcaption}
\usepackage[margin=1in]{geometry}
\usepackage{cite}
\usepackage{sectsty}
\allsectionsfont{\sffamily}
\usepackage[colorlinks,allcolors=blue]{hyperref}
\DeclareMathAlphabet{\mathscrbf}{OMS}{mdugm}{b}{n}

\let\OLDthebibliography\thebibliography
\renewcommand\thebibliography[1]{
  \OLDthebibliography{#1}
  \setlength{\parskip}{0pt}
  \setlength{\itemsep}{6.1pt plus 0.3ex}
}

\newcommand{\be}{\begin{equation}}
\newcommand{\ee}{\end{equation}}
\newcommand{\ben}{\begin{enumerate}}
\newcommand{\een}{\end{enumerate}}
\newcommand{\bi}{\begin{itemize}}
\newcommand{\ei}{\end{itemize}}
\newcommand{\bmm}{\begin{pmatrix}}
\newcommand{\emm}{\end{pmatrix}}

\newcommand{\Ad}{\text{Ad}}
\newcommand{\bra}{\langle}

\newcommand{\demi}{\frac{1}{2}}
\newcommand{\der}{\partial}
\newcommand{\Diff}{\text{Diff}\,S^1}

\newcommand{\eg}{e.g.\ }
\newcommand{\hDiff}{\widehat{\text{Diff}}\,S^1}
\newcommand{\hG}{\widehat{G}}
\newcommand{\hU}{\,\widehat{\cal U}}
\newcommand{\heta}{\widehat{\Theta}}
\newcommand{\ie}{i.e.\ }
\newcommand{\ket}{\rangle}
\newcommand{\nn}{\nonumber}
\newcommand{\phii}{\varphi}
\newcommand{\PSL}{\text{PSL}(2,\mathbb{R})}
\renewcommand{\refeq}[1]{\stackrel{\text{(\ref{#1})}}{=}}
\newcommand{\SL}{\text{SL}(2,\mathbb{R})}
\renewcommand{\sl}{\mathfrak{sl}(2,\mathbb{R})}
\newcommand{\SU}{\text{SU}(2)}

\newcommand{\un}{\text{U}(1)}
\newcommand{\Vect}{\text{Vect}\,S^1}

\newcommand{\bB}{\textbf B}
\newcommand{\bS}{\textbf S}

\newcommand{\cM}{{\cal M}}
\newcommand{\cO}{{\cal O}}

\newcommand{\cU}{\,{\cal U}}
\newcommand{\cW}{{\cal W}}

\newcommand{\mg}{\mathfrak{g}}
\newcommand{\cu}{\mathfrak{u}}

\newcommand{\sH}{\mathscr{H}}

\newcommand{\sfC}{\mathsf{C}}
\newcommand{\sfD}{\mathsf{D}}

\newcommand{\HH}{\mathbb{H}}
\newcommand{\NN}{\mathbb{N}}
\newcommand{\RR}{\mathbb{R}}
\newcommand{\ZZ}{\mathbb{Z}}

\begin{document}

\hrule
\begin{center}
\Large{\bfseries{\textsf{Berry Phases on Virasoro Orbits}}}
\end{center}
\hrule
~\\

\begin{center}
\large{\textsf{Blagoje Oblak$^{*}$}}
\end{center}
~\\

\begin{center}
\begin{minipage}{.9\textwidth}\small\it
\begin{center}
Institut f\"ur Theoretische Physik\\
ETH Z\"urich\\
CH-8093 Z\"urich, Switzerland
\end{center}
\end{minipage}
\end{center}

\vspace{2cm}

\begin{center}
\begin{minipage}{.9\textwidth}
\begin{center}{\bfseries{\textsf{Abstract}}}\end{center}
We point out that unitary representations of the Virasoro algebra contain Berry phases obtained by acting on a primary state with conformal transformations that trace a closed path on a Virasoro coadjoint orbit. These phases can be computed exactly thanks to the Maurer-Cartan form on the Virasoro group, and they persist after combining left- and right-moving sectors. Thinking of Virasoro representations as particles in AdS$_3$ dressed with boundary gravitons, the Berry phases associated with Brown-Henneaux diffeomorphisms provide a gravitational extension of Thomas precession.
\end{minipage}
\end{center}

\vfill
\noindent
\mbox{}
\raisebox{-3\baselineskip}{%
\parbox{\textwidth}{\mbox{}\hrulefill\\[-4pt]}}
{\scriptsize$^*$ E-mail: boblak@phys.ethz.ch.}

\thispagestyle{empty}
\newpage

\textsf{\tableofcontents}
\thispagestyle{empty}

\newpage
\setcounter{page}{1}
\pagenumbering{arabic}

\section{Introduction}

Berry phases commonly occur in quantum systems whose Hamiltonian depends continuously on certain external parameters \cite{Berry:1984jv,pancharatnam1956generalized}. If the parameters are fixed once and for all, any state vector with energy $E$ evolves in time according to multiplication by the dynamical phase factor $e^{-iEt/\hbar}$. But when these parameters are slowly varied, the Hamiltonian becomes time-dependent and gives rise to an extra phase, on top of the dynamical one, picked up by any energy eigenstate along time evolution. For arbitrary paths in parameter space, this phase is ambiguous; but for closed paths it becomes a well-defined, observable {\it Berry phase} that follows from the parameter-dependence of energy eigenstates. The prototypical example of a system displaying such phenomena is a spin degree of freedom coupled to a uniform magnetic field with constant norm but varying direction: the Hamiltonian is essentially the projection of the spin operator along the magnetic field, so the space of parameters can be identified with a sphere and the Berry phase picked up by a spin eigenstate along a closed curve in parameter space is proportional to the area enclosed by that path on the sphere.\\

The latter example is actually a special case of the type of systems that we wish to investigate in this paper. Namely, consider a unitary representation $\cU$ of a symmetry group $G$ that contains time translations. The Hamiltonian operator $H$ is one of the generators of the Lie algebra of $G$, but what exactly one decides to call `the Hamiltonian' relies on a choice of frame. Indeed, for any group element $f\in G$, the operator $\cU[f]H\cU[f]^{-1}$ is just as good a Hamiltonian as $H$ itself. One can think of $f$ as a change of reference frame; then $\cU[f]H\cU[f]^{-1}$ is the Hamiltonian of an observer whose frame is related by $f$ to a frame where the Hamiltonian is $H$. Thus the representation comes equipped with an entire family of Hamiltonians $\cU[f]H\cU[f]^{-1}$, each labelled by a group element $f$. Moreover, if $|\phi\ket$ is an eigensate of $H$, then $\cU[f]|\phi\ket$ is an eigenstate of $\cU[f]H\cU[f]^{-1}$ with the same eigenvalue. When considering a closed path $f(t)$ in $G$, one obtains a time-dependent Hamiltonian $\cU[f(t)]H\cU[f(t)]^{-1}$ and it is natural to investigate the Berry phase picked up by the states $\cU[f(t)]|\phi\ket$. Such symmetry-based Berry phases were studied \eg in \cite{jordan1988berry}, and it was shown in \cite{boya2001berry} that they are related to symplectic fluxes on coadjoint orbits --- an observation that we will encounter repeatedly below. The above example of a spin system corresponds to precisely such a configuration, where the Hilbert space is that of an irreducible unitary representation of $\SU$.\\

Our goal in this paper is to compute the Berry phases that occur in unitary representations of the Virasoro algebra. As we shall see, this relies on a relation between the Maurer-Cartan form on the Virasoro group and the Berry connection. The statement that Virasoro representations contain Berry phases is not new: to our knowledge it first appeared long ago in \cite{Mickelsson:1987mx} with a formula for the Berry curvature, though the actual value of these phases was not worked out. A similar curvature also appears in \cite{Bradlyn:2015wsa}, albeit in a different language and in the context of the quantum Hall effect. Furthermore the Maurer-Cartan form of Virasoro is known \cite{Alekseev:1988ce,Alekseev:1990mp,Rai:1989js}, as is the relation between Maurer-Cartan forms and Berry connections. But it seems that these coexisting patches of literature have never been properly linked, so the first purpose of this work is to put these partial results in a consistent whole and to point out that Virasoro Berry phases can be evaluated exactly, despite living on an infinite-dimensional parameter space --- see formula (\ref{ss68}) below. A second purpose is to relate these phases to gravity: since Virasoro describes the asymptotic symmetries of gravitation on three-dimensional anti-de Sitter space (AdS$_3$) \cite{Brown:1986nw}, one may ask if the Berry phases obtained here have a bulk interpretation when thinking of Virasoro representations as particles dressed with boundary gravitons. To answer this we show that the Berry phases associated with AdS$_3$ isometries reproduce known formulas for Thomas precession \cite{Thomas:1926dy,aravind1997wigner}. The phases due to asymptotic symmetry transformations that do {\it not} belong to the isometry subgroup may then be seen as new, gravitational contributions to Thomas precession. By extension to four-dimensional BMS symmetry \cite{Bondi:1960jsa}, this suggests that one can associate Berry phases with soft gravitons, or equivalently gravitational vacua \cite{Strominger:2013jfa}.\\

The plan of this paper is as follows. We start in section \ref{ss15} by reviewing Berry phases in general and explaining how they are related to Maurer-Cartan forms in the context of group representations. In section \ref{s57} we turn to the Virasoro group and evaluate its Maurer-Cartan form; the latter has already appeared in \cite{Alekseev:1988ce,Alekseev:1990mp,Rai:1989js}, but to our knowledge the derivation displayed here is new. Section \ref{s63} is the core of the paper: in it we compute explicit Berry phases associated with arbitrary closed paths in the space of conformal transforms of a primary state. By combining left and right conformal groups, these phases may be seen as observable quantities associated with boundary gravitons in AdS$_3$. Finally, section \ref{s98} is devoted to various open issues that stem from our analysis, while appendices \ref{appb} and \ref{appa} illustrate our approach with representations of $\SU$ and $\SL$.

\section{Berry phases and group representations}
\label{ss15}

In this section we briefly review general aspects of Berry phases and apply them to unitary group representations, where the Berry connection is closely related to the Maurer-Cartan form. We refer \eg to \cite[chap.\ 10]{griffiths2016introduction} and \cite[chap.\ 17]{messiah1995mecanique} or the reviews \cite{Shapere:1989kp} for an introduction to Berry phases; see also \cite{Baggio:2017aww} for recent works on Berry phases in field theory. The second part of this section will rely on various tools in group theory and differential geometry; for more on this, see \eg \cite[chap.\ 5]{sternberg1999lectures} or \cite[chap.\ 5]{Oblak:2016eij}. Note that our presentation will not be mathematically rigorous. In particular, all manifolds, bundles, functions and sections below are assumed to be smooth (except if explicitly stated otherwise).

\subsection{Generalities on Berry phases}
\label{s15}

Consider a quantum system with Hilbert space $\sH$ whose Hamiltonian depends on certain continuous external parameters. One can think of them as coordinates on a manifold $\cM$ and associate with each point $p\in\cM$ a Hamiltonian operator $H(p)$. We shall assume that the energy spectrum is discrete at all points of $\cM$; then for each $p\in\cM$ the eigenvalues $E_n(p)$ of $H(p)$ can be labelled by an integer $n\in\NN$. Each $E_n$ defines a vector bundle \cite{Simon:1983mh}
\be
\Big\{
(p,|\phi\ket)
\in\cM\times\sH
\Big|
H(p)|\phi\ket
=
E_n(p)|\phi\ket
\Big\}
\label{s1b}
\ee
whose base space is $\cM$ and whose fibre at $p$ is the subspace of $\sH$ generated by eigenvectors with eigenvalue $E_n(p)$. Let us focus on one such eigenvalue $E_n$, assuming for simplicity that it is non-degenerate for all $p\in\cM$. Then the corresponding bundle (\ref{s1b}) has one-dimensional fibres: it is a complex line bundle. If for each $p$ we let $|\psi_n(p)\ket$ be a normalized eigenvector of $H(p)$ with eigenvalue $E_n(p)$, the map $|\psi_n(\cdot)\ket:\cM\rightarrow\sH:p\mapsto|\psi_n(p)\ket$ is a section of that bundle. Note that for any function $\alpha$ on $\cM$ the vectors $e^{i\alpha(p)}|\psi_n(p)\ket$ are still normalized eigenstates of $H(p)$ with the same eigenvalues $E_n(p)$, so there are infinitely many possible choices for the section $|\psi_n(\cdot)\ket$; different sections are related by $\un$ gauge transformations on $\cM$. (For degenerate eigenvalues the gauge group is non-Abelian \cite{Wilczek:1984dh}, but we will only deal with cases where such complications do not arise.)\\

Now suppose that the system is initially prepared in the state vector $|\psi_n(p)\ket$ and let the parameters vary in time so as to trace a path $\gamma(t)$ on $\cM$ with $\gamma(0)=p$. At this point we do not ask {\it why} the parameters vary; typically, their value is fixed by an experimenter looking at the system from the outside, who can tune them at will. Thus the system is not isolated and its Hamiltonian operator $H(\gamma(t))$ changes in time. We will only assume that this change is very slow (adiabatic) in the sense that all eigenvalues of $dH(\gamma(t))/dt$ are negligible with respect to $\Delta E^2/\hbar$, where $\Delta E>0$ is the smallest energy gap between $|\psi_n\ket$ and any other energy eigenstate on the curve $\gamma$.\footnote{See \cite{Aharonov:1987gg} for an approach that does not rely on the adiabatic approximation.} The adiabatic theorem \cite{Born1928} then ensures that at any time $t$ the probability of finding the system in the state $|\psi_n(\gamma(t))\ket$ equals one; in other words the wavefunction $|\psi(t)\ket$ at time $t$ can be written as
\be
|\psi(t)\ket
=
e^{i\theta_n(t)}|\psi_n(\gamma(t))\ket+\cdots
\label{s3}
\ee
where $\theta_n$ is some real phase and the ellipsis represents terms that vanish in the adiabatic limit. By plugging this Ansatz in the Schr\"odinger equation and taking the scalar product with $|\psi_n(\gamma(t))\ket$, one finds that in the adiabatic limit
\be
\theta_n(T)
=
-\frac{1}{\hbar}\int_0^Tdt\,E_n(\gamma(t))
+
i\int_0^T\!dt\,
\big<\psi_n(\gamma(t))\big|
\frac{\der}{\der t}
\big|\psi_n(\gamma(t))\big>.
\label{s18}
\ee
Here the first term is the expected dynamical phase due to the fact that each $|\psi_n(\gamma(t))\ket$ is an energy eigenstate. But the interesting piece is the second term, which accounts for the fact that the vector $|\psi_n(p)\ket$ depends on the point $p\in\cM$ in parameter space; it is an extra phase of purely geometric (non-dynamical) origin, due to the `twisting' of the line bundle (\ref{s1b}). More precisely, one can associate with the vectors $|\psi_n(p)\ket$ a {\it Berry connection}
\be
A_n
\equiv
i\bra\psi_n(\cdot)|d|\psi_n(\cdot)\ket
\label{ss18}
\ee
where the exterior derivative $d$ is that of the parameter manifold $\cM$. By definition, this is a one-form on $\cM$ which, when paired with a vector $\dot\gamma(t)$ tangent to $\cM$ at $\gamma(t)$, returns a real number\footnote{Reality follows from $\bra\psi_n(p)|\psi_n(p)\ket=1$ $\forall$ $p\in\cM$, implying that $\bra\psi_n|d|\psi_n\ket$ is purely imaginary.}
\be
\big(A_n\big)_{\gamma(t)}(\dot\gamma(t))
=
i\big<\psi_n(\gamma(t))\big|
\frac{\der}{\der t}
\big|\psi_n(\gamma(t))\big>.
\nn
\ee
It follows that the geometric part of the phase (\ref{s18}) can be written as an integral  of the Berry connection over the path $\gamma$:
\be
\theta_{n,\text{geom}}
\equiv
i\int_0^T\!dt\,\big<\psi_n(\gamma(t))\big|
\frac{\der}{\der t}
\big|\psi_n(\gamma(t))\big>
=
\int_{\gamma}A_n.
\label{s19}
\ee
Since $A_n$ is a one-form, this expression is invariant under reparametrizations of $\gamma(t)$.\\

The terminology here stems from the fact that $A_n$ is actually a connection one-form for the line bundle (\ref{s1b}). In particular, when changing the local phase of $|\psi_n(p)\ket$ according to $|\psi_n(p)\ket\mapsto e^{i\alpha(p)}|\psi_n(p)\ket$ where $\alpha$ is some function on $\cM$, the one-form (\ref{ss18}) changes as
\be
A_n=i\bra\psi_n|d|\psi_n\ket
\mapsto
i\bra\psi_n|d|\psi_n\ket-d\alpha=A_n-d\alpha
\nn
\ee
so its transformation law is indeed that of a $\un$ gauge field (connection) over $\cM$. In that language the geometric phase (\ref{s19}) is generally gauge-dependent, since for generic $\gamma$ its value depends on the phase of the vectors $|\psi_n(p)\ket$. But for {\it closed} curves $\gamma$ the integral (\ref{s19}) becomes independent of this arbitrary choice, leading to a gauge-invariant {\it Berry phase} \cite{Berry:1984jv}
\be
B_n[\gamma]
=
\oint_{\gamma}i\bra\psi_n|d|\psi_n\ket
=
\oint_{\gamma}A_n.
\label{s20}
\ee
This expression is the (logarithm of the) holonomy of the connection (\ref{ss18}) along the path $\gamma$. Using Stokes' theorem it can be written as the flux of the Berry curvature $F_n\equiv dA_n$ through any two-surface with boundary $\gamma$. In physical terms, eq.\ (\ref{s20}) gives the geometric phase picked up by the wavefunction (\ref{s3}) between the times $t=0$ and $t=T$. It can be observed in an interference experiment involving two copies of the same system $\sH$, one with fixed parameters, the other with parameters following the curve $\gamma$. In the remainder of this section we explain how Berry phases appear in unitary group representations.

\subsection{Berry phases in group representations}
\label{ss21}

Let $G$ be a connected Lie group; we write its elements as $f$, $g$, etc.\ and let $e$ denote the identity. The tangent space $T_eG\equiv\mg$ is the Lie algebra of $G$ and we write its elements as $X$, $Y$, etc. Let also $\cU$ be a unitary representation of $G$ in some Hilbert space $\sH$; it associates with each $f\in G$ a unitary operator $\cU[f]$, so $G$ can be interpreted as the symmetry group of some quantum system. We shall assume that $G$ contains a one-parameter subgroup generated by a Lie algebra element $X_0\in\mg$ such that all group elements $e^{tX_0}$ are interpreted as `time translations'. Then $\cU[e^{tX_0}]$ is an evolution operator and the Hamiltonian is the Hermitian operator $H\equiv i\cu[X_0]$, where $\cu$ is the differential of $\cU$ at the identity. Equivalently, $\cu$ is the Lie algebra representation such that
\be
\cU[e^{tX}]=e^{t\,\cu[X]}
\quad
\text{for all }X\in\mg.
\label{s21}
\ee
Since $\cU$ is unitary, $\cu[X]$ is an anti-Hermitian operator for any $X\in\mg$.\\

The very fact that time translations are part of the symmetry group implies that the choice of Hamiltonian is generally not unique. Indeed, any group element $f$ can be seen as a `change of reference frame': if Alice measures a Hamiltonian $H$ and if Bob's frame is related to Alice by the transformation $f$, then Bob measures a Hamiltonian $\cU[f]H\cU[f]^{-1}$. One can thus associate with the group $G$ and the representation $\cU$ an entire family of Hamiltonian operators $\cU[f]H\cU[f]^{-1}$ labelled by the elements of $G$. In this picture the group manifold $G$ is a space of `parameters' $f\in G$ on which the Hamiltonian depends, and one may ask if this leads to Berry phases when $f(t)$ is a closed path in $G$.\\

To answer this question, let $|\phi\ket\in\sH$ be a normalized eigenstate of $H$ with eigenvalue $E$, which we assume to be isolated and non-degenerate. Then for each $f\in G$ the vector $\cU[f]|\phi\ket$ is a normalized eigenstate of $\cU[f]H\cU[f]^{-1}$ with the same eigenvalue. If we let $f:[0,T]\rightarrow G:t\mapsto f(t)$ be a path in $G$, the Hamiltonian $\cU[f(t)]H\cU[f(t)]^{-1}$ becomes time-dependent; provided the process is adiabatic, the initial state $\cU[f(0)]|\phi\ket$ then evolves into a state $e^{i\theta(T)}\cU[f(T)]|\phi\ket$ with the phase (\ref{s18}) now given by
\be
\theta(T)
=
-\frac{ET}{\hbar}
+
i\int_0^T\!dt\,
\big<\phi\big|
\cU[f(t)]^{\dagger}\frac{\der}{\der t}\cU[f(t)]
\big|\phi\big>.
\nn
\ee
This contains a trivial dynamical piece involving $ET$, as well as a geometric piece (\ref{s19}); the latter is due to the fact that the state $\cU[f]|\phi\ket$ depends on the choice of `frame' $f$. When the path $f(t)$ is closed, this geometric contribution becomes a Berry phase (\ref{s20}):
\be
B_{\phi}[f]=\oint_fi\bra\phi|\cU[\cdot]^{-1}d\,\cU[\cdot]|\phi\ket,
\label{s23}
\ee
where we used unitarity to write $\cU[\cdot]^{\dagger}=\cU[\cdot]^{-1}$. We denote by $\cU[\cdot]^{-1}d\,\cU[\cdot]$ the one-form on $G$ which, when paired with a vector $\dot f(t)$ tangent to $G$ at $f(t)$, returns an anti-Hermitian operator
\be
\cU[f(t)]^{-1}\frac{\der}{\der t}\cU[f(t)]
=
\cu\left[
\left.\frac{\der}{\der\tau}\right|_{\tau=t}\left(f(t)^{-1}\cdot f(\tau)\right)
\right]
\in\text{End}(\sH),
\nn
\ee
where $\cu$ is the Lie algebra representation defined by (\ref{s21}). The argument of $\cu$ may be recognized as the {\it Maurer-Cartan form}\footnote{To be precise we should refer to $\Theta$ as the {\it left} Maurer-Cartan form (it is invariant under left multiplication). There also exists a right Maurer-Cartan form, but we will not need it.} $\Theta$ on $G$ (see \eg \cite[chap.\ 5]{sternberg1999lectures} or \cite[chap.\ 5]{Oblak:2016eij}); it is the $\mg$-valued one-form on $G$ that associates with a vector $\dot f(t)\in T_{f(t)}G$ (where $f(t)$ is a path in $G$) the Lie algebra element
\be
\Theta_{f(t)}(\dot f(t))
\equiv
\left.\frac{d}{d\tau}\right|_{\tau=t}
\left[
f(t)^{-1}\cdot f(\tau)
\right].
\label{t24}
\ee
Note that here the derivative only hits on $f(\tau)$, and not on $f(t)^{-1}$. For matrix groups one can rewrite (\ref{t24}) as $f(t)^{-1}\dot f(t)$, but for more abstract groups (such as Virasoro) the general definition (\ref{t24}) is required. In any case, in terms of the Maurer-Cartan form the group-theoretic Berry phase (\ref{s23}) can be compactly written as
\be
B_{\phi}[f]
=
\oint_fi\bra\phi|\cu[\Theta]|\phi\ket
\equiv
\oint_fA_{\phi}
\label{s24}
\ee
where the Berry connection $A_{\phi}$, given in general by (\ref{ss18}), now coincides with the expectation value of the Hermitian operator $i\cu[\Theta]$ representing (\ref{t24}) in the state $|\phi\ket$:
\be
A_{\phi}
=
\bra\phi|i\cu[\Theta]|\phi\ket.
\label{ss24}
\ee
In short, the Maurer-Cartan form of a Lie group can be used to evaluate the Berry phases of its unitary representations. See appendices \ref{appb} and \ref{appa} for applications of this method to highest-weight representations of $\SU$ and $\SL$, respectively.\\

Before applying this to Virasoro, let us investigate some important geometric aspects of Berry phases in group representations. First, in deriving eq.\ (\ref{s24}) we demanded that the path $f(t)$ in $G$ be closed, \ie $f(T)=f(0)$. But this is overly restrictive: all we really need is that the vectors $\cU[f(T)]|\phi\ket$ and $\cU[f(0)]|\phi\ket$ belong to the same ray in $\sH$ so that
\be
\cU[f(T)]|\phi\ket
=
e^{i\theta}\cU[f(0)]|\phi\ket
\label{s25}
\ee
for some phase $\theta\in\RR$ (which has nothing to do with the Berry phase). How should one evaluate the Berry phase for such an open path? To answer this, let $G_{\phi}$ be the stabilizer of $|\phi\ket$, \ie the subgroup of $G$ whose elements leave invariant the ray of $|\phi\ket$:
\be
G_{\phi}
\equiv
\big\{
h\in G\,\big|\cU[h]|\phi\ket\propto|\phi\ket
\big\}.
\nn
\ee
Then eq.\ (\ref{s25}) says that $f(T)$ and $f(0)$ differ by an element of the stabilizer: $f(T)=f(0)\cdot h$ with $\cU[h]|\phi\ket=e^{i\theta}|\phi\ket$. If $h$ is not the identity, the curve $f(t)$ is open and formula (\ref{s24}) does not apply; this can be corrected by defining a closed path
\be
\bar f(t)
=
\begin{cases}
f(t) & \text{for }t\in[0,T]\\
f(T)\cdot h(t) & \text{for }t\in[T,T']
\end{cases}
\label{path}
\ee
where $h(t)$ is a curve in $G_{\phi}$ such that $h(T)=e$ and $h(T')=h^{-1}=f(T)^{-1}f(0)$.\footnote{Here we are assuming that $h$ belongs to the component of the identity in $G_{\phi}$.} Since $\bar f(t)$ is closed, it can be plugged in eq.\ (\ref{s24}); the result is a sum of two terms coming from the two pieces of (\ref{path}): a bulk term that coincides with (\ref{s24}) but with open $f(t)$, and a boundary term due to $h(t)$ that cancels the spurious phase $\theta$ of (\ref{s25}). That boundary term is independent of the choice of path $h(t)$, since it only depends on its endpoints, so from now on we define Berry phases for {\it open} paths $f$ by $B_{\phi}[f]\equiv B_{\phi}[\bar f]$:
\be
\boxed{B_{\phi}[f]
=
\int_fi\big<\phi\big|
\cu[\Theta]
\big|\phi\big>
-i\log\big<\phi\big|
\cU[f(0)^{-1}f(T)]
\big|\phi\big>.}
\label{s26}
\ee
When $f(T)=f(0)$, the logarithmic boundary term vanishes and $B_{\phi}[f]$ reduces to (\ref{s24}).\\

Note the change of perspective of the last few lines: from now on, we no longer need to assume that the path $f(t)$ is closed, as long as its endpoint $f(T)$ is such that $f(0)^{-1}f(T)$ belongs to the stabilizer of $|\phi\ket$. To further illustrate this point, suppose that $f(t)$ is a curve (not necessarily closed) entirely contained within $G_{\phi}$ so that $\cU[f(t)]|\phi\ket=e^{i\theta(t)}|\phi\ket$ for all $t$. Then $f(0)^{-1}f(T)$ certainly belongs to $G_{\phi}$, so eq.\ (\ref{s26}) applies and one can verify that it leads to a vanishing Berry phase. This confirms that the relative sign in (\ref{s26}) is correct: for $f(t)\in G_{\phi}$, each $f(t)$ leaves the ray of $|\phi\ket$ unaffected and adiabatic evolution along such a path should have no physical effect, so the corresponding Berry phase must vanish. More generally, any curve of the form $f(t)=f(0)\cdot h(t)$ with $h(t)\in G_{\phi}$ gives a vanishing Berry phase.

\subsection{Coadjoint orbits and quantization}
\label{s27}

The fact that curves contained in the stabilizer of $|\phi\ket$ have zero Berry phases illustrates an important point: when `varying the parameters' of the representation by adiabatically following a path $f(t)$ in $G$, what truly matters for the Berry phase is not the path itself, but rather its projection down to the quotient space $G/G_{\phi}=\{f\cdot G_{\phi}|f\in G\}$. The phase can be non-zero only when this {\it projected} path is non-trivial. From that point of view, allowing the curve $f(t)$ in (\ref{s26}) to be open provided $f(0)^{-1}f(T)\in G_{\phi}$ is just saying that the corresponding projected path in $G/G_{\phi}$ must be closed. Thus the Berry phase (\ref{s26}) is related to the geometry of the group manifold $G$, seen as a principal $G_{\phi}$-bundle over the quotient space $G/G_{\phi}$. In particular the Berry connection that gives rise to the phase is not quite the one in (\ref{ss24}), which lives on the group manifold, but rather its pullback to $G/G_{\phi}$ by a section of the bundle. The boundary term in eq.\ (\ref{s26}) accounts for this subtlety.\\

These observations are related to geometric quantization and the `orbit method' for building group representations \cite{kirillov2004lectures,Witten:1987ty}: one can think of $G/G_{\phi}$ as a homogeneous space that coincides with a coadjoint orbit of $G$ when $|\phi\ket$ is a coherent state corresponding to the `highest weight point' on the orbit.\footnote{For the definition of coherent states in group representations, see \eg \cite{ali2000coherent}.} Then the pullback of (\ref{ss24}) by a section of the bundle $G\rightarrow G/G_{\phi}$ coincides with the symplectic potential on $G/G_{\phi}$ and the Berry curvature coincides with the natural symplectic form on the orbit (the {\it Kirillov-Kostant} symplectic form). In particular, the Berry phase (\ref{s26}) is the flux of the symplectic form through any surface whose boundary is the projection of $f(t)$ on $G/G_{\phi}$ \cite{boya2001berry}. Our goal in this paper is to show that such fluxes can be evaluated {\it exactly} on Virasoro coadjoint orbits, despite their being infinite-dimensional (see \cite{neeb2002central} for related considerations).

\begin{figure}[t]
\centering
\includegraphics[width=0.50\textwidth]{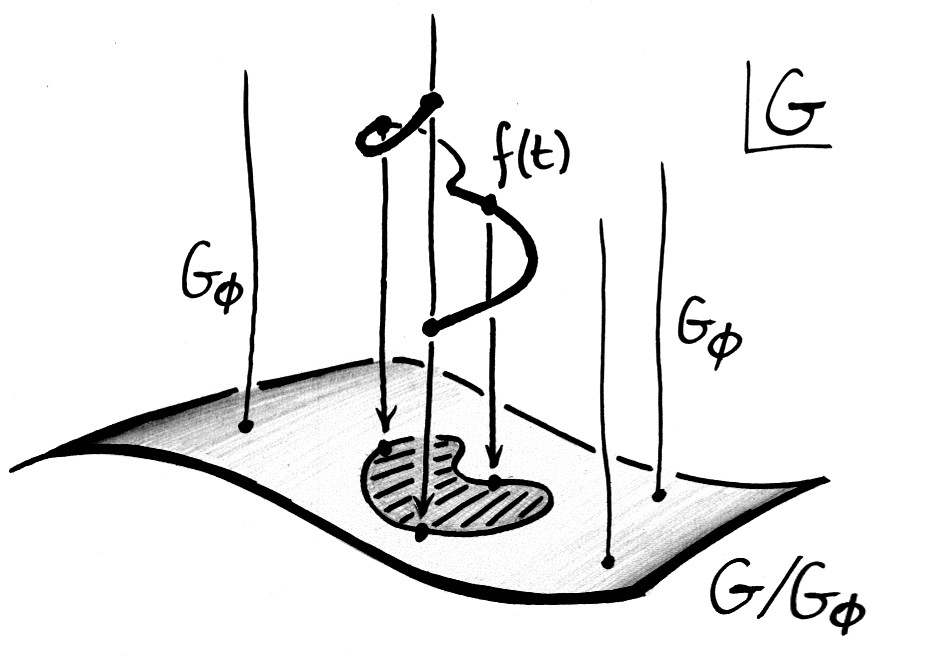}
\caption{A path $f(t)$ in $G$ such that $f(T)=f(0)\cdot h$ with $h\in G_{\phi}$ is mapped on a closed curve in $G/G_{\phi}$ by the projection $\pi:G\rightarrow G/G_{\phi}:f\mapsto f\cdot G_{\phi}$. When $G/G_{\phi}$ is a coadjoint orbit, the Berry phase along $f(t)$ coincides with the flux of the Kirillov-Kostant symplectic form through any surface enclosed by the projected path on $G/G_{\phi}$. For $f(t)=f(0)\cdot h(t)$ \mbox{with $h(t)\in G_{\phi}$, the projected path is a point and the Berry phase vanishes.}\label{s31}}
\end{figure}

\section{Maurer-Cartan form on the Virasoro group}
\label{s57}

In this section we describe the Virasoro group, \ie the central extension of the group of diffeomorphisms of the circle, and evaluate its Maurer-Cartan form (\ref{t24}). This is a necessary first step towards the Virasoro Berry phases of section \ref{s63}. The computation relies on certain group-theoretic tools that we will describe only superficially; we refer \eg to \cite[chap.\ 4]{guieu2007algebre} or \cite[chap.\ 6]{Oblak:2016eij} for a much more thorough presentation.

\subsection{Virasoro group}
\label{t57}

Let $\Diff$ be the group of diffeomorphisms of the circle with the group operation given by composition of maps; writing its elements as $f$, $g$, etc., their product is  $f\cdot g\equiv f\circ g$. $\Diff$ is an infinite-dimensional Lie group with two connected components; the identity component is the group $\text{Diff}^+S^1$ of orientation-preserving diffeomorphisms. The group elements are most easily described by thinking of $S^1$ as a quotient $\RR/2\pi\ZZ$ and labelling its points by an angular coordinate $\phii\in\RR$ identified as $\phii\sim\phii+2\pi$. In these terms, any element of $\text{Diff}^+S^1$ can be represented by a smooth map $f:\RR\rightarrow\RR:\phii\mapsto f(\phii)$ such that
\be
f'(\phii)>0
\quad\text{and}\quad
f(\phii+2\pi)=f(\phii)+2\pi
\qquad\forall\,\phii\in\RR.
\label{ss57}
\ee
The identity is $e(\phii)=\phii$ and the inverse $f^{-1}$ of $f$ is such that $f(f^{-1}(\phii))=f^{-1}(f(\phii))=\phii$.\\

The set of maps satisfying (\ref{ss57}) is actually the universal covering group $\widetilde{\text{Diff}}{}^+S^1$, while $\text{Diff}^+S^1$ as such is obtained by enforcing the extra identification $f\sim f+2\pi$. (The proper terminology is that functions satisfying (\ref{ss57}) are {\it lifts} of $S^1$ diffeomorphisms, but we will simply refer to them as diffeomorphisms of the circle.) If one thinks of $\phii$ as a light-cone coordinate $x^+$ say, then the group $\widetilde{\text{Diff}}{}^+S^1$ of maps $x^+\mapsto f(x^+)$ is the chiral half of the conformal group of a Lorentzian cylinder. In that sense $\widetilde{\text{Diff}}{}^+S^1$ is a group of conformal transformations. To lighten the notation, from now on we simply denote $\widetilde{\text{Diff}}{}^+S^1$ by $\Diff$ and take all its elements to be maps $\phii\mapsto f(\phii)$ which satisfy (\ref{ss57}).\\

The Virasoro group is the universal central extension of $\Diff$. It can be described as follows: for any two elements $f,g\in\Diff$, we define a real number
\be
\sfC(f,g)
\equiv
-\frac{1}{48\pi}
\int_0^{2\pi}\!d\phii\,\log\big[f'(g(\phii))\big]\frac{g''(\phii)}{g'(\phii)}.
\label{s58}
\ee
This specifies a map $\sfC:\Diff\times\Diff\rightarrow\RR$ known as the {\it Bott(-Thurston) cocycle} \cite{bott1977characteristic}. Then the {\it Virasoro group} is defined as the set
\be
\hDiff
\equiv
\Diff\times\RR
\label{vigo}
\ee
whose elements are pairs $(f,\alpha)$ with $f\in\Diff$ and $\alpha\in\RR$, multiplied according to
\be
(f,\alpha)\cdot(g,\beta)
\equiv
\big(
f\circ g,\alpha+\beta+\sfC(f,g)
\big).
\label{s59}
\ee
One may verify that this is indeed a group operation; for instance, associativity follows from the cocycle identity satisfied by (\ref{s58}):
\be
\sfC(f,g)+\sfC(f\circ g,h)
=
\sfC(f,g\circ h)+\sfC(g,h).
\nn
\ee
Using $\sfC(f,f^{-1})=\sfC(f^{-1},f)=0$, the product (\ref{s59}) implies that the inverse of $(f,\alpha)$ is
\be
(f,\alpha)^{-1}
=
(f^{-1},-\alpha).
\label{ivi}
\ee
We will use this below to evaluate the Maurer-Cartan form (\ref{t24}).\\

At this point, the definition of the Virasoro group, and in particular the Bott cocycle (\ref{s58}), seem to come completely out of the blue. We will not go into the details of their construction here; instead we will verify below that the Lie algebra of the group defined here is the familiar Virasoro algebra. Furthermore one can show that the Bott cocycle measures the symplectic area of certain triangles on Virasoro coadjoint orbits \cite{neeb2002central}; this observation is closely related to Berry phases, and indeed we shall see that these phases are essentially integrals of the differential of the Bott cocycle.

\subsection{Virasoro algebra}
\label{seval}

For future reference it is useful to review the link between the Virasoro group and its Lie algebra. First recall that the flow of any vector field defines a one-parameter family of diffeomorphisms, so the Lie algebra of $\Diff$ is the space $\Vect$ of vector fields on the circle; we write them as $X=X(\phii)\frac{\der}{\der\phii}$ with a $2\pi$-periodic component $X(\phii)$. Similarly, the Lie algebra of the Virasoro group (\ref{vigo}) is $\Vect\oplus\RR$, whose elements are pairs $(X,\alpha)$ consisting of a vector field $X$ and a real number $\alpha$. It remains to find the Lie bracket of these elements, which we will do by evaluating the differential of the adjoint representation of the Virasoro group. This procedure is described in greater detail in \cite[sec.\ 6.4]{Oblak:2016eij} (see also \cite[app.\ A.1]{Afshar:2015wjm} for similar considerations in the warped Virasoro group).\\

By definition, the adjoint representation of a Lie group $G$ is the map that associates with $f\in G$ a linear operator $\Ad_f$ acting on the Lie algebra $\mg$ according to
\be
\Ad_f(X)
=
\left.\frac{d}{dt}\right|_{t=0}f\cdot e^{tX}\cdot f^{-1}
\qquad\forall\,X\in\mg.
\nn
\ee
For matrix groups this reduces to $\Ad_fX=fXf^{-1}$, but this is not true in general. For $\Diff$, the group operation is given by composition and one can show that the adjoint representation coincides with the transformation law of vector fields on the circle:
\be
\big(\Ad_fX\big)(\phii)
=
\frac{X(f^{-1}(\phii))}{(f^{-1})'(\phii)},
\quad\text{\ie}\quad
\big(\Ad_fX\big)(f(\phii))
=
f'(\phii)X(\phii).
\label{adif}
\ee
Here both sides of the equations should be interpreted as the component of a vector field. Of course the story is more complicated when the central extension is switched on: in that case one must use the group operation (\ref{s59}) and the inverse (\ref{ivi}) to evaluate
\be
(f,\alpha)\cdot(e^{tX},t\beta)\cdot(f,\alpha)^{-1}
=
\Big(
f\circ e^{tX}\circ f^{-1},
t\beta+\sfC\big(f,e^{tX}\big)+\sfC\big(f\circ e^{tX},f^{-1}\big)
\Big)
\nn
\ee
where $\sfC$ is the Bott cocycle (\ref{s58}). The first entry of this expression is the one that leads to (\ref{adif}), but the second one is new; one can show that the result is
\be
\Ad_{(f,\alpha)}(X,\beta)
=
\Big(
\Ad_fX,
\beta-\frac{1}{24\pi}\int_0^{2\pi}\!d\phii\,\{f;\phii\}X(\phii)
\Big)
\label{adiv}
\ee
where the first entry is the centreless adjoint representation (\ref{adif}) while the second entry involves the Schwarzian derivative
\be
\{f;\phii\}
=
\frac{f'''}{f'}-\frac{3}{2}\left(\frac{f''}{f'}\right)^2
\label{sadi}
\ee
with the right-hand side evaluated at $\phii$. Note that in (\ref{adiv}) the central element $\alpha$ acts trivially, as it should. So much for the adjoint representation of the Virasoro group.\\

By differentiating the adjoint representation of a group, one can read off the bracket of its Lie algebra. Thus we define the brackets of elements of the Virasoro algebra as
\be
\big[
(X,\alpha),(Y,\beta)
\big]
\equiv
\left.\frac{d}{dt}\right|_{t=0}\Ad_{(e^{tX},t\alpha)}(Y,\beta).
\label{adad}
\ee
Using the Virasoro adjoint representation (\ref{adiv}), one then finds
\be
\big[
(X,\alpha),(Y,\beta)
\big]
=
\Big(\!-\![X,Y],-\frac{1}{24\pi}\int_0^{2\pi}\!d\phii\,X'''(\phii)Y(\phii)\Big)
\label{liba}
\ee
where the first entry on the right-hand side is the {\it opposite} of the usual Lie bracket of vector fields,
\be
-[X(\phii)\der_{\phii},Y(\phii)\der_{\phii}]
=
\big(\!-\!X(\phii)Y'(\phii)+Y(\phii)X'(\phii)\big)\der_{\phii}.
\nn
\ee
This awkward sign is actually common for groups of diffeomorphisms (see \eg \cite[exercise 4.1G]{abraham1978foundations}); it can be absorbed by adding an extra sign in the definition of the bracket (\ref{adad}), but here we will stick to the sign in (\ref{liba}) since it allows us to use the same conventions for Berry phases as in section \ref{ss21}. As for the second entry of (\ref{liba}), it is a central extension that involves a term $X'''$ due to the Schwarzian derivative (\ref{sadi}) of $f(\phii)=\phii+tX(\phii)+\cO(t^2)$ \cite{gel1968cohomologies}. Note that the central terms $\alpha$, $\beta$ in (\ref{liba}) do not contribute to the bracket, as they should.\\

To relate eq.\ (\ref{liba}) to the standard presentation of the Virasoro algebra in physics, recall that $X(\phii)$ and $Y(\phii)$ are functions on the circle and can be expanded in Fourier modes. This motivates the definition of the (complex) Virasoro generators
\be
\ell_m\equiv\Big(-ie^{im\phii}\der_{\phii},-\frac{i}{24}\delta_{m,0}\Big),
\quad
Z\equiv(0,-i),
\qquad m\in\ZZ.
\label{vigen}
\ee
Any element of the Virasoro algebra can then be written as a linear combination $(X,\alpha)=\sum_{m\in\ZZ}X_m\ell_m+i\alpha Z$ with $X_m^*=-X_{-m}$. The bracket (\ref{liba}) yields
\be
[\ell_m,\ell_n]
=
(m-n)\ell_{m+n}+\frac{Z}{12}m(m^2-1)\delta_{m+n,0}
\label{vicom}
\ee
while all brackets involving $Z$ vanish. Thus $Z$ is a central charge that commutes with all Virasoro generators, and takes a definite value $Z=c$ in any (irreducible) representation of the Virasoro algebra. We will use these properties in section \ref{s63} below.

\subsection{Maurer-Cartan form}

We now evaluate the Maurer-Cartan form (\ref{t24}) on the Virasoro group (\ref{vigo}). To do this we will first deal with the centreless group $\Diff$, then include its central extension (\ref{s58}). The Maurer-Cartan form for Virasoro has previously appeared in \cite{Alekseev:1988ce,Alekseev:1990mp,Rai:1989js}, without reference to the group operation (\ref{s59}). Our derivation here will be technically different in that it relies heavily on the Bott cocycle (\ref{s58}), but the result will of course be identical.\\

To begin, consider the group $\Diff$ of (lifts of orientation-preserving) diffeomorphisms of the circle. At any point $f\in\Diff$, the tangent space $T_f\Diff$ consists of time derivatives of paths $\gamma:(-\epsilon,\epsilon)\rightarrow\Diff:t\mapsto\gamma(t,\cdot)$ such that $\gamma(0,\phii)=f(\phii)$. Then the Maurer-Cartan form (\ref{t24}) evaluated at $f$ is a linear isomorphism mapping $T_f\Diff$ on $T_e\Diff=\Vect$. To evaluate it, we let $\gamma$ be a path in $\Diff$ such that $\gamma(0)=f$ and use the chain rule to act with the Maurer-Cartan form $\Theta$ on the vector $\dot\gamma(0)$:\footnote{The dot in $\dot\gamma$ denotes a time derivative, while the middle dot in (\ref{s61}) denotes standard pointwise multiplication of functions on the circle.}
\be
\Theta_f\big(\dot\gamma(0)\big)
\refeq{t24}
\frac{d}{dt}\big(f^{-1}\circ\gamma(t)\big)
=
\dot\gamma(0)\cdot\big(f^{-1}\big)'\circ f.
\label{s61}
\ee
Note that the function $\dot\gamma(0,\phii)$ is $2\pi$-periodic since each $\gamma(t,\phii)$ is a diffeomorphism that satisfies (\ref{ss57}), so (\ref{s61}) is a vector field $X(\phii)\frac{\der}{\der\phii}$ on $S^1$ whose $2\pi$-periodic component is
\be
X(\phii)
=
\Theta_f\big(\dot\gamma(0)\big)(\phii)
\refeq{s61}
\dot\gamma(0,\phii)\cdot\big(f^{-1}\big)'(f(\phii))
=
\frac{\dot\gamma(0,\phii)}{f'(\phii)},
\label{ss61}
\ee
where we used $f'\cdot(f^{-1})'\circ f=1$. This defines the Maurer-Cartan form on the group $\Diff$ and is sometimes written as $\Theta_f=\delta f/f'$ \cite{Alekseev:1988ce}.\\

Let us now apply similar arguments to the Virasoro group. To lighten the notation, we write $G=\Diff$ and $\hG=\hDiff=G\times\RR$; the group elements are pairs $(f,\alpha)$, so a path in the Virasoro group is written as $(\gamma(t),\mu(t))$ with $\gamma$ a path in $G$ and $\mu$ a path on the real line. If $\gamma(0)=f$ and $\mu(0)=\alpha$, this path defines a vector $(\dot\gamma(0),\dot\mu(0))$ tangent to $\hG$ at the point $(f,\alpha)$. The Maurer-Cartan form (\ref{t24}) evaluated at $(f,\alpha)$ maps this vector on a certain element of the Virasoro algebra, which we want to find. Denoting the Maurer-Cartan form on the Virasoro group by $\heta$, the definition (\ref{t24}) gives
\begin{align}
\heta_{(f,\alpha)}\big(\dot\gamma(0),\dot\mu(0)\big)
&=
\left.\frac{d}{dt}\right|_{t=0}\Big[(f,\alpha)^{-1}\cdot\big(\gamma(t),\mu(t)\big)\Big]\nn\\
\label{maca}
&=
\left.\frac{d}{dt}\right|_{t=0}\Big(f^{-1}\circ\gamma(t),-\alpha+\mu(t)+\sfC\big(f^{-1},\gamma(t)\big)\Big)
\end{align}
where we used the inverse (\ref{ivi}) and the group operation (\ref{s59}). The time derivative applies separately to the two entries of this expression. In the first entry we recognize the centreless Maurer-Cartan form $\Theta$ given by (\ref{s61}), so we can write
\be
\heta_{(f,\alpha)}\big(\dot\gamma(0),\dot\mu(0)\big)
=
\Big(
\Theta_f\big(\dot\gamma(0)\big),
\dot\mu(0)+\left.\frac{d}{dt}\right|_{t=0}\sfC\big(f^{-1},\gamma(t)\big)
\Big)
\label{macabo}
\ee
where the time-independent term $\alpha$ of (\ref{maca}) does not contribute. We can make the structure of this result more apparent by defining for each $f\in G$ a linear map
\be
\sfD_f:T_fG\rightarrow\RR:\dot\gamma(0)\mapsto\sfD_f\big(\dot\gamma(0)\big)=\left.\frac{d}{dt}\right|_{t=0}\sfC\big(f^{-1},\gamma(t)\big),
\label{ss62}
\ee
which is essentially the derivative of the cocycle $\sfC$ with respect to its second argument (hence the notation `$\sfD$'). Thinking of the tangent space $T_{(f,\alpha)}\hG$ as a direct sum $T_fG\oplus T_{\alpha}\RR=T_fG\oplus\RR$ and similarly for the Lie algebra $T_{(e,0)}\hG=T_eG\oplus\RR$, the Maurer-Cartan form (\ref{macabo}) is a linear map
\be
\heta_{(f,\alpha)}:
T_{(f,\alpha)}\hG\rightarrow T_{(e,0)}\hG:
\bmm \dot\gamma(0) \\ \dot\mu(0) \emm
\mapsto
\heta_{(f,\alpha)}\cdot\bmm \dot\gamma(0) \\ \dot\mu(0) \emm
\nn
\ee
that can be written as a matrix
\be
\heta_{(f,\alpha)}
=
\bmm
\Theta_f & 0 \\
\sfD_f & 1
\emm.
\label{s62}
\ee
Here $\Theta$ is the centreless Maurer-Cartan form on $G$ (given by (\ref{ss61}) for $G=\Diff$) and $\sfD$ is the map (\ref{ss62}); were it not for $\sfD$, the centrally extended Maurer-Cartan form (\ref{s62}) would be a diagonal map. In particular, when the central path $\mu(t)$ is set to zero, the action of (\ref{s62}) on the tangent vector $(\dot\gamma(0),0)$ reduces to
\be
\heta_{(f,\alpha)}\big(\dot\gamma(0),0\big)
=
\Big(
\Theta_f\big(\dot\gamma(0)\big),
\sfD_f\big(\dot\gamma(0)\big)
\Big).
\label{labok}
\ee
Note that these results hold for any Lie group with a differentiable central extension $\sfC$.\\

As far as the Maurer-Cartan form of the Virasoro group is concerned, we already know its centreless piece (\ref{ss61}); in order to write the matrix (\ref{s62}) explicitly we only have to compute the operator $\sfD_f$ defined in (\ref{ss62}). So let $\gamma(t)$ be a path in $\Diff$ such that $\gamma(0)=f$; it defines a tangent vector $\dot\gamma(0)\in T_f\Diff$. Using the Bott cocycle (\ref{s58}) and an integration by parts, one finds
\be
\sfD_f\big(\dot\gamma(0)\big)
=
-\frac{1}{48\pi}\int_{S^1}\left[
\left.\frac{\der}{\der t}\right|_0\Big(\log\big((f^{-1})'\circ\gamma(t)\big)\Big)
+
\left.\frac{\der}{\der t}\right|_0\log(\gamma(t)')
\right]d\log(f').
\nn
\ee
Performing the time derivatives and integrating by parts once more to remove the $\phii$ derivative from $\dot\gamma'(0,\phii)$, one ends up with
\be
\sfD_f\big(\dot\gamma(0)\big)
=
\frac{1}{48\pi}\int_0^{2\pi}\!d\phii\,\frac{\dot\gamma(0,\phii)}{f'(\phii)}\left(\frac{f''}{f'}\right)'(\phii).
\label{sad}
\ee
This is a linear functional of $\dot\gamma(0,\cdot)$, as it should. Combined with the centreless Maurer-Cartan form (\ref{ss61}), it entirely specifies the Maurer-Cartan form (\ref{s62}) of the Virasoro group.\\

For future reference, let us rewrite (\ref{ss61}) and (\ref{sad}) with a simpler notation. Namely, instead of denoting the path on $\Diff$ by $\gamma(t,\phii)$, we write it as $f(t,\phii)$ where each $f(t,\cdot)$ is a diffeomorphism of the circle.\footnote{Sometimes we also write $f(t,\cdot)\equiv f(t)$ provided there is no danger of confusion.} Furthermore, at each time $t$ the time derivative $\dot f(t,\cdot)$ is a vector tangent to $\Diff$ at $f(t,\cdot)$. Then, using (\ref{labok}) and the results (\ref{ss61}) and (\ref{sad}), we can write the action of the Virasoro Maurer-Cartan form $\heta$ on $\dot f(t,\cdot)$ as
\be
\heta_{(f,\alpha)}\big(\dot f(t,\cdot),0\big)
=
\Bigg(
\frac{\dot f(t,\phii)}{f'(t,\phii)}\frac{\der}{\der\phii}\,,\,
\frac{1}{48\pi}\int_0^{2\pi}\!d\phii\,\frac{\dot f(t,\phii)}{f'(t,\phii)}\left(\frac{f''(t,\phii)}{f'(t,\phii)}\right)'
\Bigg).
\label{s61b}
\ee
The entries on the right-hand side are respectively a vector field on the circle and a real number, so the result belongs to the Virasoro algebra as it should. It contains a non-zero central piece due to the Bott cocycle. It now remains to use this formula to evaluate the Berry phases of Virasoro representations.

\section{Virasoro Berry phases}
\label{s63}

In this section we evaluate the Berry phases (\ref{s26}) that appear in unitary highest weight representations of the Virasoro algebra. After deriving the general formula, which will coincide with the geometric action functional studied in \cite{Alekseev:1988ce,Alekseev:1990mp,Rai:1989js}, we apply it to circular paths in the space of conformal transformations. When combining left- and right-moving sectors and identifying the conformal group with the asymptotic symmetry group of AdS$_3$ gravity, our results generalize Thomas precession. We also briefly discuss the relation of these observations with gravitational memory.

\subsection{General derivation}
\label{gede}

Consider a unitary highest weight representation $\cu$ of the Virasoro algebra with central charge $c>0$ and highest weight $h$. The generators (\ref{vigen}) are represented by operators
\be
\cu[\ell_m]\equiv L_m
\quad\text{and}\quad
\cu[Z]=c\hat{I},
\qquad m\in\ZZ,
\nn
\ee
where $\hat{I}$ is the identity. Unitarity means that any real vector field $X(\phii)\der_{\phii}$ is represented by an anti-Hermitian operator $\cu[X]$, which implies that the operators $L_m$ satisfy the Hermiticity conditions $L_m^{\dagger}=L_{-m}$. Their commutators reproduce the bracket (\ref{vicom}):
\be
[L_m,L_n]
=
(m-n)L_{m+n}+\frac{c}{12}m(m^2-1)\delta_{m+n,0}.
\nn
\ee
A basis of the Hilbert space $\sH$ is provided by the highest weight state $|h\ket$ such that
\be
L_0|h\ket=h|h\ket,
\quad
L_m|h\ket=0\text{ if }m>0,
\label{hiwe}
\ee
together with its descendants $L_{-m_1}...L_{-m_N}|h\ket$, where $1\leq m_1\leq...\leq m_N$. For the vacuum representation $h=0$ we also demand $\SL$ invariance, $L_{-1}|0\ket=0$.\\

We will assume that the algebra representation $\cu$ integrates into a representation $\hU$ of the Virasoro group; the hat stresses that $\hU$ represents Virasoro rather than just $\Diff$. Since $\cu$ is the differential of $\hU$ at the identity, eq.\ (\ref{s21}) can now be written as
\be
\hU[(e^{tX},t\alpha)]=e^{t\,\cu[(X,\alpha)]}
\quad
\text{for all }(X,\alpha)\in\Vect\oplus\RR.
\nn
\ee
The representation $\hU$ is such that $\hU[(f,\alpha)]=e^{ic\,\alpha}\cU[f]$, where $\cU[f]$ is a unitary operator acting on $\sH$. Compatibility with the group operation (\ref{s59}) implies that the operators $\cU[f]$ satisfy the composition law
\be
\cU[f]\circ\cU[g]
=
e^{ic\,\sfC(f,g)}\cU[f\circ g]
\nn
\ee
where $\sfC$ is the Bott cocycle (\ref{s58}). Thus $\cU$ represents the group $\Diff$ up to a generally non-vanishing phase due to the central extension (for more on representations up to phases, see \eg \cite[sec.\ 2.2]{Weinberg:1995mt} or \cite[chap.\ 2]{Oblak:2016eij}).\\

As before, we are interested in the Berry phases that appear when applying symmetry transformations to an energy eigenstate. Specifically, consider the highest weight vector $|h\ket$; it is an eigenstate of the `standard Hamiltonian' $L_0$. For $h>0$ the stabilizer of $|h\ket$ is the $\un$ subgroup of $\Diff$ generated by $L_0$, but generic conformal transformations act non-trivially on $|h\ket$: the set of physically inequivalent states that can be reached from $|h\ket$ is an infinite-dimensional homogeneous manifold $\Diff/S^1$.\footnote{More precisely, the $\un$ subgroup of $\text{Diff}{}^+S^1$ becomes an $\RR$ subgroup of the universal covering $\widetilde{\text{Diff}}{}^+S^1$ such that $\text{Diff}{}^+S^1/S^1=\widetilde{\text{Diff}}{}^+S^1/\RR$; we write this as $\Diff/S^1$ for brevity.} It is in fact a coadjoint orbit of the Virasoro group \cite{Lazutkin,Witten:1987ty}, and to evaluate Berry phases we will need to compute certain holonomies along closed paths on that orbit. In principle we could choose a section of the bundle $\Diff\rightarrow\Diff/S^1$ and use it to pullback the Maurer-Cartan form, but in practice it will be much easier to use eq.\ (\ref{s26}) directly. In that approach our weapon of choice will be the Maurer-Cartan form (\ref{s61b}). Similar considerations apply to the vacuum case $h=0$, except that the corresponding orbit is $\Diff/\PSL$ due to the enhanced stabilizer.\\

With the notation used above (\ref{s61b}), let $f(t,\cdot)$ be a path in $\Diff$; for each $t\in[0,T]$ we have a circle diffeomorphism $\phii\mapsto f(t,\phii)$.\footnote{Technically we should consider a path $(f(t),\mu(t))$ in the Virasoro group with a central piece $\mu(t)$, but the latter does not contribute to Berry phases so we set it to zero without loss of generality.} In order for the curve to be closed when projected on $\Diff/S^1$, we require that $f(0)^{-1}\circ f(T)$ be a rotation by some angle $\theta$:
\be
f^{-1}\big(0,f(T,\phii)\big)=\phii+\theta.
\label{t66}
\ee
Then eq.\ (\ref{s26}) says that we can associate with that path a well-defined Berry phase
\be
B_{h,c}[f]
=
\int_fi\bra h|\cu[\heta]|h\ket
-i\log\bra h|\hU\Big[\big(f(0)^{-1},0\big)\cdot\big(f(T),0\big)\Big]|h\ket
\label{s66}
\ee
picked up by the states $\hU\big[(f(t),0)\big]|h\ket$ along the curve $f(t)$. Here $\heta$ is the Maurer-Cartan form (\ref{s62}) and contains two terms: a centreless piece $\Theta$ that coincides with (\ref{ss61}), and a central piece $\sfD$ written in (\ref{sad}). When plugged into (\ref{s66}), the centreless piece yields
\be
\int_fi\bra h|\cu\big[(\Theta,0)\big]|h\ket
=
\int_0^T\!dt
\,i\bra h|
\cu\bigg[
\bigg(
\frac{\dot f}{f'},0
\bigg)
\bigg]
|h\ket.
\label{ss66}
\ee
Since $\dot f/f'$ is the component of a vector field (\ref{ss61}) on the circle, it can be written as a Fourier series with modes $e^{im\phii}$. When plugged into $\cu$, the $m^{\text{th}}$ mode gives a contribution proportional to the operator $L_m$. But to evaluate (\ref{ss66}) we only need the expectation value in the state $|h\ket$ and eq.\ (\ref{hiwe}) implies $\bra h|L_m|h\ket=0$ for $m\neq 0$, so the only mode that survives is the zeroth one. The latter is the integral of $\dot f/f'$ over $S^1$ and one has
\be
\int_fi\bra h|\cu\big[(\Theta,0)\big]|h\ket
=
-\left(h-\frac{c}{24}\right)
\frac{1}{2\pi}
\int_0^T\!dt\int_0^{2\pi}\!d\phii\,\frac{\dot f(t,\phii)}{f'(t,\phii)},
\label{s67}
\ee
where the coefficient $h-c/24$ comes from $L_0|h\ket=h|h\ket$ and the normalization of $\ell_0$ in (\ref{vigen}). So much for the centreless piece of (\ref{s66}). For the central piece we recall that $\cu[(0,\alpha)]|\phi\ket=ic\,\alpha|\phi\ket$ for any vector $|\phi\ket$, so we can write
\be
\int_fi\bra h|\cu[(0,\sfD)]|h\ket
=
-c\int_0^T\!dt\,\sfD_f(\dot f)
\refeq{sad}
-\frac{c}{48\pi}\int_0^T\!dt\int_0^{2\pi}\!d\phii\,\frac{\dot f(t,\phii)}{f'(t,\phii)}\left(\frac{f''(t,\phii)}{f'(t,\phii)}\right)'.
\label{ss67}
\ee
We can now combine (\ref{s67}) and (\ref{ss67}) to write the complete Berry phase (\ref{s66}) as
\be
B_{h,c}[f]
=
-\frac{1}{2\pi}
\int_0^T\!\!dt\int_0^{2\pi}\!\!d\phii\,
\frac{\dot f}{f'}\left[
h-\frac{c}{24}
+
\frac{c}{24}\left(\frac{f''}{f'}\right)'\,
\right]
-i\log\bra h|\cU\big[f(0)^{-1}\circ f(T)\big]|h\ket.
\label{s68}
\ee
To simplify this further, recall that $f(0)^{-1}\circ f(T)$ is a rotation (\ref{t66}) by an angle $\theta=f^{-1}\big(0,f(T,0)\big)$. Since the state $|h\ket$ is an eigenvector of $L_0$, a rotation by $\theta$ maps it on
\be
\hU[(\text{rot}_{\theta},0)]|h\ket
=
e^{i\theta(L_0-c/24)}|h\ket
=
e^{i\theta(h-c/24)}|h\ket.
\nn
\ee
It follows that the Berry phase (\ref{s68}) can finally be rewritten as
\be
\boxed{%
B_{h,c}[f]
=
-\frac{1}{2\pi}
\int_0^T\!dt\int_0^{2\pi}\!d\phii\,
\frac{\dot f}{f'}\left[
h-\frac{c}{24}
+
\frac{c}{24}\left(\frac{f''}{f'}\right)'\,
\right]
+\left(h-\frac{c}{24}\right)
f^{-1}\big(0,f(T,0)\big).}
\label{ss68}
\ee
This is our main result; the phase is clearly non-zero for generic paths in $\Diff$, and it is manifestly invariant under reparametrizations of $t$ (as it should). Furthermore the formula is explicit: in principle, given any time-dependent family of diffeomorphisms $f(t,\phii)$, one can plug it in (\ref{ss68}) and read off the Berry phase. The latter can also be seen as the holonomy of a Berry connection on an infinite-dimensional manifold $\Diff/S^1$, with a curvature that coincides with the Kirillov-Kostant symplectic form \cite{Mickelsson:1987mx}. In line with the general results outlined in section \ref{s27}, the Berry phase can thus be interpreted as the flux of the symplectic form through any surface whose boundary is the path $f(t,\cdot)$ projected on $\Diff/S^1$ (or $\Diff/\PSL$ if $h=0$). This generalizes the formulas of appendix \ref{appa} to the Virasoro group and allows us to interpret the Berry phase as (the kinetic piece of) an action functional. Indeed, eq.\ (\ref{ss68}) coincides with the `geometric action' describing the path $f(t,\phii)$ when the latter is seen as a dynamical field on a cylinder \cite{Alekseev:1988ce,Alekseev:1990mp,Rai:1989js};\footnote{See also \cite{Manvelyan:1993np} for related considerations, and \cite{Mandal:2017thl} for a recent application to the SYK model.} for comparison we refer \eg to eq.\ (25) of \cite{Alekseev:1988ce}, noting that what the authors call $F$ in that reference is what we would write as $f^{-1}$ and that their $b_0$ is our $(h-c/24)/2\pi$.

\subsection{Circular paths and superboosts}
\label{cipa}

We now apply formula (\ref{ss68}) to curves $f(t)$ such that the expectation value of $L_0$ is constant in the states $\cU[f(t)]|h\ket$. Those are circular paths, each of which can be written as
\be
f(t,\phii)
=
g(\phii)+\omega t
\label{s71}
\ee
where $g$ is some fixed diffeomorphism and $\omega>0$ is some angular velocity. In other words $f(t)=\text{rot}_{\omega t}\circ g$, the entire time-dependence being contained in the rotation. The time parameter ranges from $t=0$ to some upper bound $t=T$; we take $T=2\pi/\omega$ so that (\ref{s71}) turns exactly once around the tip of $\Diff/S^1$. The corresponding diffeomorphism (\ref{t66}) is a pure rotation
\be
f^{-1}\big(
0,f(2\pi/\omega,\phii)
\big)
=
g^{-1}\big(
g(\phii)+2\pi
\big)
\refeq{ss57}
\phii+2\pi,
\nn
\ee
so the Berry phase formula (\ref{ss68}) applies. Note that this relies crucially on the value of $T$: only for $T=2\pi n/\omega$ with integer $n$ is it true that $f^{-1}\big(0,f(T,\phii)\big)=\phii+2\pi n$ regardless of $g$; in general, only this discrete family of intervals gives closed paths on $\Diff/S^1$. (Certain non-generic choices of $g$ may be `more symmetric' and ensure that $f^{-1}\big(0,f(T,\phii)\big)$ is a rotation even for lower values of $T$; this will be the case for superboosts below.)\\

Let us now write the Berry phase (\ref{ss68}) for the circular path (\ref{s71}) and $T=2\pi/\omega$. Since $\dot f=\omega$ and $f'=g'$, the computation of the time integral is straightforward and gives
\be
B_{h,c}[\text{rot}_{\omega t}\circ g]
=
-\int_0^{2\pi}\frac{d\phii}{g'}\left[
h-\frac{c}{24}+\frac{c}{24}\left(\frac{g''}{g'}\right)'\,
\right]
+2\pi\left(h-\frac{c}{24}\right).
\label{s73}
\ee
This result is independent of $\omega$: the Berry phase is the same for any strictly positive value of angular velocity, however small. (For negative values of $\omega$ one would obtain the {\it opposite} of (\ref{s73}); more on that in section \ref{secombi}.) This is a consequence of reparametrization invariance, but one should keep in mind that Berry phases can only be observed in the adiabatic regime. In the case at hand this means that $\omega$ should be much smaller than the energy gap between $|h\ket$ and any other state; in AdS$_3$ this gap is of order $1/\ell$ in terms of the AdS$_3$ radius $\ell$, so the angular velocity must satisfy $\omega\ll1/\ell$.\\

For circular paths there is no way to go beyond eq.\ (\ref{s73}) without specifying $g(\phii)$. In what follows we evaluate (\ref{s73}) for transformations $g(\phii)$ that are `superboosts' and relate the result to Thomas precession. As a motivation, note that the Virasoro group contains infinitely many different subgroups that are all locally isomorphic to $\SL$. Indeed, let $\alpha$, $\beta$ be complex numbers such that $|\alpha|^2-|\beta|^2=1$ and let $n$ be a strictly positive integer; then one can define a diffeomorphism $f\in\Diff$ (satisfying (\ref{ss57})) by
\be
e^{inf(\phii)}
=
\frac{\alpha e^{in\phii}+\beta}{\beta^*e^{in\phii}+\alpha^*}.
\label{s74}
\ee
The set of such $f$'s (at fixed $n$ but varying $\alpha,\beta$) is a group isomorphic to the $n$-fold cover of $\text{PSL}(2,\RR)=\SL/\ZZ_2$; the $\PSL$ group leaving the vacuum invariant has $n=1$.\\

One can think of the Virasoro group as half of the asymptotic symmetry group of AdS$_3$ gravity \cite{Brown:1986nw}. A map $f\in\Diff$ then acts on the left light-cone coordinate $x^+$ on the cylinder at infinity according to $x^+\mapsto f(x^+)$. From that point of view, $\PSL$ isometries of global AdS$_3$ take the form (\ref{s74}) with $n=1$, while transformations with $n>1$ are genuine {\it asymptotic} symmetries that would not occur if gravitation was switched off. Similarly, asymptotic symmetries of Minkowskian space-times span the BMS group \cite{Bondi:1960jsa} where Poincar\'e isometries are extended to supertranslations and superrotations. To apply an analogous terminology to Virasoro, note that a map (\ref{s74}) with $n=1$ and $\alpha=\text{cosh}(\lambda/2)$, $\beta=\text{sinh}(\lambda/2)$ corresponds to (half of) an AdS$_3$ boost with rapidity $\lambda$ in the direction $\phii=0$. Accordingly we shall refer to any diffeomorphism $g$ given by
\be
e^{ing(\phii)}
=
\frac{\text{cosh}(\lambda/2)e^{in\phii}+\text{sinh}(\lambda/2)}{\text{sinh}(\lambda/2)e^{in\phii}+\text{cosh}(\lambda/2)}
\label{s75}
\ee
as a {\it superboost} of order $n$ with rapidity $\lambda$. The normalization $\lambda/2$ has to do with the fact that $\SL$ is a double cover of the group $\PSL$ spanned by transformations (\ref{s74}).\\

\begin{figure}[t]
\centering
\begin{subfigure}{0.20\textwidth}
  \centering
  \includegraphics[width=\linewidth]{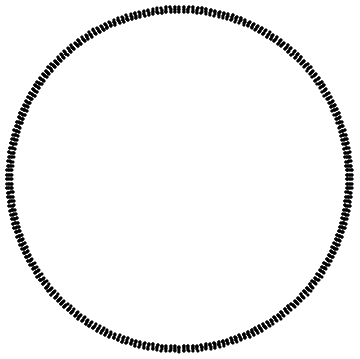}
\end{subfigure}%
$\quad\xrightarrow{~~~~~~~~~}\quad$%
\begin{subfigure}{0.20\textwidth}
  \centering
  \includegraphics[width=\linewidth]{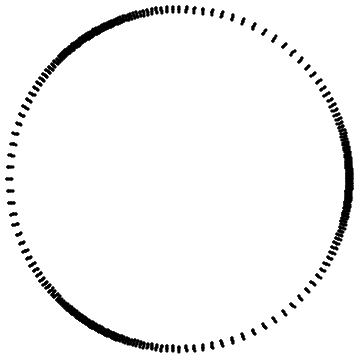}
\end{subfigure}
\caption{\label{supabo} A circle with uniformly spaced dots is acted upon by a superboost (\ref{s75}) of order $n=3$ with positive $\lambda$. The dots converge towards the points $\phii=0,2\pi/3,4\pi/3$ and move away from $\phii=\pi/3,\pi,5\pi/3$. Analogous observations apply to all superboosts.}
\end{figure}

Our goal now is to evaluate the Berry phase (\ref{s73}) for a circular path (\ref{s71}) in which $g$ is a superboost. Thanks to the $\ZZ_n$ symmetry of (\ref{s75}) under $\phii\mapsto\phii+2\pi/n$, we could let the time parameter $t$ range from $0$ to $2\pi/(\omega n)$ and still get a well-defined Berry phase (\ref{ss68}); nevertheless we will use the same time interval $[0,2\pi/\omega]$ as in the general case. Then, using (\ref{s75}) to evaluate derivatives of $g$, the phase (\ref{s73}) reads
\be
B_{h,c}[\text{rot}_{\omega t}\circ g]
=
-2\pi\left(h-\frac{c}{24}\right)\big(\cosh\lambda-1\big)
-\frac{c\,n^2\text{sinh}\lambda}{24}
\int_0^{2\pi}\!dx\,
\frac{\sin^2x}{\text{cotanh}\lambda+\cos x},
\nn
\ee
where we have changed variables according to $x=n\phii$. It remains to evaluate the integral over $x$; this can be done for instance thanks to the residue theorem, and yields\footnote{With a time interval $[0,2\pi/(\omega n)]$ the result (\ref{t78}) would be divided by $n$.}
\be
\boxed{\bigg.
B_{h,c}(n,\lambda)
=
-2\pi\left(h+\frac{c}{24}(n^2-1)\right)
\big(\text{cosh}\lambda-1\big).}
\label{t78}
\ee
This is the Berry phase associated with a circular path of superboosts of order $n$ and with rapidity $\lambda$; one can view it as the integral over a disk of the Berry curvature displayed in eq.\ (3.11) of \cite{Mickelsson:1987mx} (see also eq.\ (5.11) of \cite{Bradlyn:2015wsa}). It vanishes for $\lambda=0$, as it should since in that case the transformation (\ref{s75}) is the identity. Note also that for $n=1$ the term proportional to the central charge disappears and (\ref{t78}) reduces to an $\SL$ Berry phase,\footnote{See appendix \ref{appa} for the derivation of (\ref{s53}) in a highest weight representation of $\SL$.}
\be
B_{h,c}(1,\lambda)
=
-2\pi h(\cosh\lambda-1).
\label{s53}
\ee
This formula should be familiar from special relativity: when $h$ is identified with the spin of a particle that follows a circular trajectory at rapidity $\lambda$, the phase (\ref{s53}) reflects the net rotation undergone by its locally inertial reference frame after one revolution. This phenomenon is known as {\it Thomas precession} \cite{Thomas:1926dy}; it is due to the fact that products of non-collinear Lorentz boosts contain Wigner rotations \cite{Wigner:1939cj} (see also \cite[chap.\ 6]{Misner:1974qy} or \cite[sec.\ 11.8]{Jackson:1998nia}). In this sense Virasoro Berry phases include and extend Thomas precession.

\subsection{Combining left and right sectors}
\label{secombi}

So far we have only considered one chiral copy of the Virasoro group; we now combine two of them into a direct product, as follows. Consider a two-dimensional conformal field theory (CFT) on a Lorentzian cylinder with radius $\ell$ and a metric proportional to
\be
ds^2
=
-dt^2+\ell^2d\phii^2
\label{ss78}
\ee
where $t\in\RR$ is a time coordinate and $\phii\in\RR$ is an angular coordinate identified as $\phii\sim\phii+2\pi$. These can be combined into dimensionless light-cone coordinates
\be
x^{\pm}
\equiv
\frac{t}{\ell}\pm\phii.
\label{s78}
\ee
In terms of $x^{\pm}$, conformal transformations of the cylinder take the form
\be
(x^+,x^-)
\mapsto
\big(
f(x^+),\bar f(x^-)
\big)
\label{ss79}
\ee
where $f$ and $\bar f$ are independent diffeomorphisms of the real line that preserve the orientation of the cylinder in the sense that $f'(x^+)$ and $\bar f'(x^-)$ are positive. Moreover they must preserve the identification $\phii\sim\phii+2\pi$, which implies that $f$ and $\bar f$ both satisfy the conditions (\ref{ss57}) in terms of their arguments. It follows that the group of orientation-preserving conformal transformations of a Lorentzian cylinder is a direct product $\widetilde{\text{Diff}}{}^+S^1\times\widetilde{\text{Diff}}{}^+S^1$ whose elements are pairs $(f,\bar f)$ acting according to (\ref{ss79}). As in section \ref{t57}, we will lighten the notation by writing that group simply as $\Diff\times\Diff$.\\

It was shown in \cite{Brown:1986nw} that gravity on AdS$_3$ admits fall-off conditions such that the resulting asymptotic symmetry group coincides with the conformal group of a Lorentzian cylinder. In that picture the length scale $\ell$ in (\ref{ss78}) coincides with the AdS$_3$ radius and the light-cone coordinates (\ref{s78}) label points at spatial infinity, \ie on the cylindrical boundary of AdS$_3$. Asymptotic symmetry transformations are pairs $(f,\bar f)\in\Diff\times\Diff$ that act on $x^{\pm}$ according to (\ref{ss79}) up to corrections that vanish at infinity. In addition, the realization of the asymptotic symmetry group on phase space is projective, with non-zero left and right central charges whose standard normalization is $c=\bar c=3\ell/2G$ ($G$ being the Newton constant in three dimensions). In parity-breaking theories of gravity such as topologically massive gravity \cite{Deser:1981wh}, essentially the same conclusions hold up to the fact that left and right central charges may differ (see \eg \cite{Henneaux:2009pw}).\\

Accordingly, from now on we will think of the product of two Virasoro groups as the asymptotic symmetry group of AdS$_3$ gravity. Irreducible unitary representations of $\hDiff\times\hDiff$ with highest weights $h,\bar h$ can then be interpreted as particles in AdS$_3$ dressed with quantized boundary gravitons, with mass $(h+\bar h)/\ell$ and spin $h-\bar h$.\footnote{The AdS$_3$ mass may receive corrections of order $1/\ell^2$, but they will be unimportant for us.} For the sake of generality we allow the left and right central charges $c,\bar c$ to differ. One may then consider the state $|h,\bar h\ket=|h\ket\otimes|\bar h\ket$ in the tensor product of two Virasoro representations, act on it with a family of transformations $\cU[f(t)]\otimes\,\bar{\cal{U}}[\bar f(t)]$, and investigate the Berry phase that arises when the path $\big(f(t),\bar f(t)\big)$ is closed up to an element of the stabilizer of $|h,\bar h\ket$. The analysis is the same as in section \ref{gede}; the result is a sum of left-and right-moving Berry phases, each of which takes the form (\ref{ss68}) with the integral over $\phii$ replaced by an integral over $x^+$ or $x^-$, respectively: 
\be
B_{h,\bar h,c,\bar c}[f,\bar f]
=
B_{h,c}[f]+B_{\bar h,\bar c}[\bar f].
\label{s83}
\ee
One should keep in mind that there is, in general, no relation between $f(t,x^+)$ and $\bar f(t,x^-)$. In particular, one is free to leave one of the two paths at the identity, say $\bar f(t,x^-)=x^-$ for all $t$, in which case (\ref{s83}) reduces to the chiral Berry phase (\ref{ss68}). Note also that the time parameter $t$ along the path $(f(t),\bar f(t))$ has a very different status than the coordinates $x^{\pm}$: when writing \eg $f(t,x^+)$, each $f(t,\cdot)$ is a chiral conformal transformation, so the dependence of $f$ on $x^+$ is merely a reminder of the fact that $f$ acts on a cylinder. On the other hand the parameter $t$ in $f(t,\cdot)$ indicates a genuine, explicit time-dependence of $f$. As before we only assume that this dependence is adiabatic.\\

Let us now consider circular paths in $\Diff\times\Diff$. In contrast to the chiral case studied earlier, we now have two possibilities for what `circular' means: a first choice is $f(t,x^+)=g(x^+)+\omega t$ and $\bar f(t,x^-)=\bar g(x^-)+\omega t$, but an inequivalent second choice is obtained by changing the relative sign in $\bar f$ while keeping the same $f$.\footnote{We could even let $f$ and $\bar f$ have different angular velocities, but we will not consider such paths here.} To illustrate the difference between those two options we can relate $x^{\pm}$ to coordinates $(t,\phii)$ thanks to eq.\ (\ref{s78}); the transformations of $(t,\phii)$ corresponding to the two cases then are
\be
\begin{array}{ll}
(t,\phii)\mapsto\big(\frac{\ell}{2}(g(x^+)+\bar g(x^-))+\ell\omega t,\demi(g(x^+)-\bar g(x^-))\big) & \text{(option 1),}\\[.2cm]
(t,\phii)\mapsto\big(\frac{\ell}{2}(g(x^+)+\bar g(x^-)),\demi(g(x^+)-\bar g(x^-))+\omega t\big) & \text{(option 2).}
\end{array}
\label{s86}
\ee
This shows that only the {\it second} choice can truly be interpreted as a `circular' or `rotating' path; by contrast, the first option is a `time-translating' path. In any case, provided $\omega>0$ is small enough in the sense that $\omega\ll1/\ell$, the time-dependence of the Hamiltonian $\cU[f(t)]L_0\cU[f(t)]^{-1}+\bar{\cal{U}}[\bar f(t)]\bar L_0\bar{\cal{U}}[\bar f(t)]^{-1}$ is adiabatic and it makes sense to investigate the Berry phase picked up by the states $\cU[f(t)]|h\ket\otimes\bar{\cal{U}}[\bar f(t)]|\bar h\ket$ as $t$ runs from $t=0$ to $t=T=2\pi/\omega$. The phase can be evaluated as in the chiral case (\ref{s73}) and reads
\be
\begin{split}
B_{h,\bar h,c,\bar c}[\text{rot}_{\omega t}\circ g,\text{rot}_{\pm\omega t}\circ\bar g]
=&
-\int_0^{2\pi}\frac{dx^+}{g'}\left[
h-\frac{c}{24}+\frac{c}{24}\left(\frac{g''}{g'}\right)'\,
\right]
+2\pi\left(h-\frac{c}{24}\right)\\
& \mp
\int_0^{2\pi}\frac{dx^-}{\bar g'}\left[
\bar h-\frac{\bar c}{24}+\frac{\bar c}{24}\left(\frac{\bar g''}{\bar g'}\right)'\,
\right]
\pm
2\pi\left(\bar h-\frac{\bar c}{24}\right)
\end{split}
\nn
\ee
where the upper and lower signs respectively correspond to the first and second options in (\ref{s86}), while prime denotes differentiation with respect to the appropriate argument. For instance, when $g$ and $\bar g$ are superboosts (\ref{s75}), the Berry phase becomes
\be
B_{h,\bar h,c,\bar c}(n,\lambda,\bar n,\bar\lambda)
=
-2\pi\!\left(h+\frac{c}{24}(n^2-1)\right)
\!\big(\text{cosh}\lambda-1\big)
\mp
2\pi\!\left(\bar h+\frac{\bar c}{24}(\bar n^2-1)\right)
\!\big(\text{cosh}\bar\lambda-1\big)
\label{ss88}
\ee
where $(\lambda,n)$ and $(\bar\lambda,\bar n)$ are the respective parameters of $g$ and $\bar g$.

\subsection{Thomas precession, boundary gravitons and memory}
\label{setome}

When $n=\bar n=1$ in eq.\ (\ref{ss88}), the contributions involving central charges vanish and the Virasoro Berry phase reduces to a sum/difference of $\SL$ Berry phases (\ref{s53}). When in addition $\lambda=\bar\lambda$, and if we choose a rotating path to select the lower sign in (\ref{ss88}), the Berry phase coincides with formula (\ref{s53}) up to the replacement of $h$ by the spin $h-\bar h$. Thus Virasoro Berry phases in AdS$_3$ contain Thomas precession \cite{Thomas:1926dy}, since (\ref{s53}) is precisely the angle of rotation undergone after one revolution by the locally inertial frame of a particle with spin $h$ moving along a circle at rapidity $\lambda$.\\

From this perspective the general formulas (\ref{ss68}) or (\ref{ss88}) may be seen as generalizations of Thomas precession corresponding to closed paths in the group of asymptotic symmetry transformations in AdS$_3$. To get a grasp on what this means, let us see how far the analogy goes. Standard Thomas precession takes place when a particle follows a closed path in momentum space (hence also in position space) as a result of some external influence --- for instance the Coulomb attraction due to an atomic nucleus. The actual shape of the path is determined by the dynamics of that interaction; for example, the velocity of a particle orbiting around a nucleus depends on their masses and electric charges. Such parameters have to be plugged into the Berry phase/Thomas precession formula and cannot be determined by group theory alone. However, once the path is known, the corresponding Berry phase is entirely determined by symmetry considerations and does {\it not} explicitly depend on any dynamical parameter. The same reasoning applies to the AdS$_3$ Berry phases (\ref{s83})-(\ref{ss88}): the one-parameter family of conformal transformations $(f(t),\bar f(t))$ corresponds to a closed path in the space of CFT stress tensors dual to a dressed particle in AdS$_3$. The actual shape of that path depends on some dynamics that need to be used as input; for instance, a field may propagate in AdS$_3$ in such a way that the back-reacted space-time metric undergoes a family of Brown-Henneaux transformations $(f(t),\bar f(t))$.\footnote{The gravitational field in three dimensions has no local degrees of freedom, so the perturbation must be caused by some other field --- \eg a scalar field or a gauge field.} If this perturbation is such that the gravitational field returns to its original configuration once the disturbance has passed, then the particle's wavefunction should pick up a Berry phase (\ref{s83}). The latter is entirely determined by Virasoro symmetry: it depends on the path $(f(t),\bar f(t))$, the weights $h,\bar h$ and the central charges $c,\bar c$, but it does {\it not} depend explicitly on the various interactions that triggered the path. This is pleasant, but it also raises the question how, concretely, one is supposed to switch on perturbations that generate gravitational Berry phases. We will not address this issue here, only briefly returning to it in the next section.\\

\begin{figure}[t]
\centering
\includegraphics[width=0.30\textwidth]{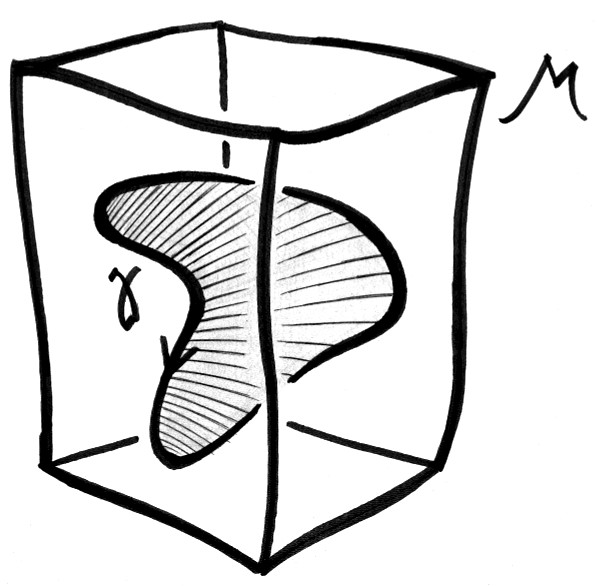}
\caption{Boundary gravitons in AdS$_3$ span an infinite-dimensional manifold $\cM\cong\Diff/\PSL\times\Diff/\PSL$. Any loop $\gamma$ in $\cM$ encloses a two-dimensional surface embedded in $\cM$. The symplectic flux through that surface coincides with the Berry phase picked up by boundary gravitons as they undergo a family of conformal transformations whose projection on $\cM$ is the path $\gamma$.}
\end{figure}

A surprising aspect of the Berry phases (\ref{s83})-(\ref{ss88}) is that they are generally {\it non-zero} even when $h=\bar h=0$ and $c=\bar c$. This means that the AdS$_3$ vacuum state $|0\ket$ can pick up a non-zero Berry phase when subjected to a one-parameter family of conformal transformations $(f(t),\bar f(t))$. In some very symmetric cases the phase vanishes (\eg when $f(t,x^+)=g(x^+)+\omega t$ and $\bar f(t,x^-)=\bar g(x^-)-\omega t$ with $g(\phii)=\bar g(\phii)$), but for generic, left-right asymmetric paths, it does not. One may think of it as a phase due to boundary gravitons dressing the AdS$_3$ vacuum. In the language of BMS symmetry \cite{Strominger:2013jfa,Strominger:2014pwa}, such phases are caused by the infinitely many vacua of the gravitational field.\footnote{The word `vacuum' here is used merely as an analogy: in contrast to BMS supertranslations, Brown-Henneaux diffeomorphisms typically change the energy of a state.} Berry phases probe the geometry of this space of vacua by measuring fluxes of its symplectic form.\\

These observations hint at a relation between Berry phases and the gravitational memory effect \cite{zel1974radiation}, which refers to the space-time displacement of observers at infinity who are exposed to a burst of gravitational radiation. It was recently shown \cite{Strominger:2014pwa} that this phenomenon can be interpreted as a BMS supertranslation caused by the passing gravitational wave; this is similar in spirit to the mechanism described above for generating Berry phases of boundary gravitons, since in the latter case we also referred to an asymptotic symmetry transformation caused by some propagating local field. Thus one can think of Virasoro Berry phases as a `memory' of dressed particles. The interpretation of Berry phases as a memory effect actually predates their very discovery \cite{Smith59} (see also \cite{berry1990anticipations}); in that respect it is no surprise that Berry phases associated with asymptotic symmetries are related to gravitational memory.

\section{Discussion}
\label{s98}

In the previous pages we have seen how Virasoro symmetry can be used to evaluate the Berry phase picked up by a primary state as it undergoes a family of conformal transformations. The result is displayed in eq.\ (\ref{ss68}) and solely depends on group-theoretic data; it can be interpreted as a flux of the natural symplectic form on an infinite-dimensional coadjoint orbit of the Virasoro group. One can also combine left and right conformal groups into a direct product, leading to a Berry phase (\ref{s83}) picked up by the wavefunction of a dressed particle in AdS$_3$ subjected to a family of Brown-Henneaux diffeomorphisms. When these transformations are just AdS$_3$ boosts, the phase reproduces Thomas precession; generic asymptotic symmetry transformations, on the other hand, provide a gravitational generalization of Thomas precession that can be interpreted as a memory effect. When applied to conformal transformations of the AdS$_3$ vacuum state, Virasoro Berry phases probe the geometry of the space of `boundary gravitons' that dress AdS$_3$.\\

These statements leave open a great many questions, both conceptual and practical. Perhaps the most pressing one is the issue raised at the end of section \ref{setome}: what dynamical mechanism, if any, could switch on Virasoro Berry phases? Since this requires an explicitly time-dependent Hamiltonian, one presumably needs to couple a two-dimensional CFT to an `external' system that triggers time-dependent changes of conformal frames. In the case of Thomas precession, the rotation of a particle around an atomic nucleus corresponds to a continuous family of Lorentz boosts with varying direction, and the source of that variation is the Coulombian attraction. To generate Virasoro Berry phases, one similarly needs to drive a CFT with some external force that obliges the stress tensor to violate its conservation law $\der_-T_{++}=\der_+T_{--}=0$. It would be interesting to find explicit field-theoretic realizations of this situation. A related problem is to understand if/how Virasoro Berry phases can be recast as statements on CFT correlation functions, and of course whether such phases have any chance of being observed in experiments. A promising setting might be provided by the quantum Hall effect, where Berry phases on an infinite-dimensional parameter space were studied in \cite{Bradlyn:2015wsa}, though the exact relation between this work and ours is not clear at this stage.\\

From the AdS$_3$ perspective, driven CFTs might be obtained by coupling gravity to a local field whose fall-off conditions allow information to flow into or out of the bulk. This kind of behaviour is admittedly more suggestive of Minkowskian physics (see \eg the Einstein-Maxwell system of \cite{Barnich:2015jua}), but surprisingly similar observations in AdS$_3$ have recently appeared in \cite{Mao:2016bzy}.\\

Aside from these obvious puzzles, the tools developed in this paper should apply to many other contexts that involve some sort of Virasoro symmetry. This includes most notably BMS in three and four dimensions, warped Virasoro \cite{Detournay:2012pc}, or the $\cW$ algebras that describe asymptotic symmetries of higher-spin theories in three dimensions \cite{Henneaux:2010xg}. We hope to say more about some of these topics in the near future.

\section*{Acknowledgements}
\addcontentsline{toc}{section}{Acknowledgements}

I am grateful to G.~Barnich, S.~Datta and M.~Gaberdiel for stimulating discussions, to M.~Kay for a short but helpful airport chat, and especially to K.~H.~Neeb for pointing me to flux homomorphisms and their relation to group cohomology, which triggered many of the ideas developed in this paper. This work is supported by the Swiss National Science Foundation, and partly by the NCCR SwissMAP.

\appendix

\section{Spin in a magnetic field}
\label{appb}

In this appendix we illustrate the method of section \ref{ss15} with the group $\SU$, which provides the textbook example of Berry phases (see \eg \cite[chap.\ 10]{griffiths2016introduction} or \cite[chap.\ 10]{Nakahara:2003nw}). Accordingly, consider an irreducible unitary representation $\cU$ of $\SU$. In physical terms this corresponds to a single (`irreducible') quantum (`unitary') spin degree of freedom (`representation of $\SU$') embedded in a uniform magnetic field $\bB$; the Hamiltonian of the system is proportional to the projection of the spin operator $\hat\bS$ along $\bB$:
\be
H=-\gamma\,\hat\bS\cdot\bB.
\label{s33}
\ee
Here $\gamma\in\RR$ is the gyromagnetic ratio and $\hat\bS=(\hat S{}_1, \hat S{}_2, \hat S{}_3)$ is a set of three Hermitian operators representing rotation generators in the Hilbert space of the representation. Their commutators are $[\hat S{}_i,\hat S{}_j]=i\epsilon_{ijk}\hat S{}_k$ (we set $\hbar=1$) and they can be written as
\be
\hat S{}_j=i\cu[-i\sigma_j/2]
\label{ss33}
\ee
where the $\sigma_j$'s are Pauli matrices and the algebra representation $\cu$ is related to $\cU$ by (\ref{s21}).\\

The hat in $\hat\bS$ stresses that the $\hat S{}_j$'s are operators, while $\bB$ is merely a set of three real parameters on which the Hamiltonian (\ref{s33}) depends. Suppose then that the magnetic field varies adiabatically by changing its direction while keeping a constant norm. From a group-theoretic standpoint this is precisely the setting described in section \ref{ss21}: in a `standard' reference frame the magnetic field is aligned with the vertical axis and the Hamiltonian (\ref{s33}) is proportional to $\hat S{}_3$, but under a rotation $f\in\SU$ the Hamiltonian becomes proportional to $\cU[f]\hat S{}_3\cU[f]^{-1}$. For a time-dependent family of rotations $f(t)$ whose projection on the sphere $\SU/S^1\cong S^2$ is a closed path, one gets a Berry phase (\ref{s26}) that we shall now compute.\\

Let $j\geq0$ be the spin of the representation $\cU$; then a basis of the carrier space of $\cU$ is provided by $(2j+1)$ states $|m\ket$ such that $\hat S{}_3|m\ket=m|m\ket$ for $m=j,j-1,...,-j$. Consider in particular the highest weight state $|j\ket$; its stabilizer is the $\un$ subgroup of $\SU$ generated by $\hat S{}_3$, but generic rotations act non-trivially on it: in terms of polar coordinates $(\theta,\phii)$ on $\SU/S^1\cong S^2$, a rotation $f$ that maps the North pole of $S^2$ on $(\theta,\phii)$ also maps $|j\ket$ on a vector $\cU[f]|j\ket$ that represents a spin pointing in the direction $(\theta,\phii)$. This provides a family of states $|j;\theta,\phii\ket$ obtained by acting on $|j\ket$ with suitable rotations; each such vector is an eigenstate of the Hamiltonian (\ref{s33}) when the magnetic field points in the direction $(\theta,\phii)$. Our goal is to find the Berry phase picked up by these states when the direction of the magnetic field traces a closed path $(\theta(t),\phii(t))$ on $S^2$. (The existence of this phase was pointed out already in Berry's seminal paper \cite{Berry:1984jv}, and its classical analogue was experimentally observed shortly thereafter \cite{Chiao:1986np}; see also \cite{leek2007observation} for recent observations of $\SU$ Berry phases in qubit systems.)\\

To describe the states $|j;\theta,\phii\ket$ we need a family of rotations $g_{(\theta,\phii)}\in\SU$, each mapping the North pole to the point $(\theta,\phii)$ on $S^2$, such that
\be
|j;\theta,\phii\ket
=
\cU[g_{(\theta,\phii)}]|j\ket.
\label{s36}
\ee
One can think of this family as a section of the $\un$ bundle $\SU\rightarrow S^2$, associating a group element $g_{(\theta,\phii)}$ with a point $(\theta,\phii)$. Since the bundle is non-trivial, there is no section that depends continuously on $(\theta,\phii)$ at all points of $S^2$; but such sections do exist locally, and this will be enough to evaluate Berry phases. In particular we may choose
\be
g_{(\theta,\phii)}
=
\bmm
\cos(\theta/2) & -e^{-i\phii}\sin(\theta/2)\\
e^{i\phii}\sin(\theta/2) & \cos(\theta/2)
\emm,
\label{s37b}
\ee
which is smooth on $S^2\backslash\{\text{South pole}\}$, since it is well-defined everywhere except at $\theta=\pi$. (Another section would have different singularities, but it would certainly have at least one due to the non-triviality of the bundle; this is an incarnation of the `Dirac strings' that arise in the study of magnetic monopoles \cite{Dirac:1931kp}.) The normalization is such that the $\text{SO}(3)$ rotation which corresponds to (\ref{s37b}) by the isomorphism $\text{SO}(3)\cong\SU/\ZZ_2$ maps the North pole of $S^2$ on the point with coordinates $(\theta,\phii)$. Indeed, with the conventions of \cite{Oblak:2015qia}, this isomorphism is obtained by defining a map
\be
F:\SU\rightarrow\text{SO}(3):\bmm a & b \\ c & d \emm\mapsto F\left[\bmm a & b \\ c & d \emm\right]
\nn
\ee
that associates an $\text{SO}(3)$ rotation with any $\SU$ group element, according to
\be
F\left[\bmm a & b \\ c & d \emm\right]
=
\bmm
\mathrm{Re}(a^*d+b^*c) & \mathrm{Im}(ad^*-bc^*) & \mathrm{Re}(a^*c-b^*d) \\
\mathrm{Im}(a^*d+b^*c) & \mathrm{Re}(ad^*-bc^*) & \mathrm{Im}(a^*c-b^*d) \\
\mathrm{Re}(ab^*-cd^*) & \mathrm{Im}(ab^*-cd^*) & \demi(|a|^2-|b|^2-|c|^2+|d|^2)
\emm.
\label{momo}
\ee
One can show that $F$ is a surjective homomorphism whose kernel consists of the identity matrix and its opposite (see \eg \cite{HenneauxGroupe,Oblak:2015qia}), which implies that $\SU$ is indeed the double cover of $\text{SO}(3)$. Applying (\ref{momo}) to the $\SU$ group element (\ref{s37b}), one finds
\be
F[g_{(\theta,\phii)}]
=
\bmm
\times & \times & \sin\theta\cos\phii\\
\times & \times & \sin\theta\sin\phii\\
\times & \times & \cos\theta
\emm
\nn
\ee
where the crossed entries are irrelevant for our purposes: the last column is enough to prove that $F[g_{(\theta,\phii)}]$ maps the North pole of $S^2$ on the point with polar coordinates $(\theta,\phii)$.\\

The section (\ref{s37b}) can be used to pullback the Maurer-Cartan form (\ref{t24}) from $\SU$ to the sphere, which results in an $\SU$ gauge field
\be
g_{(\theta,\phii)}^{-1}dg_{(\theta,\phii)}
=
\frac{i}{2}
\bmm
(1-\cos\theta)d\phii & e^{-i\phii}(\sin\theta\,d\phii+id\theta)\\
e^{i\phii}(\sin\theta\,d\phii-id\theta) & -(1-\cos\theta)d\phii
\emm
\nn
\ee
that can be rewritten as the following linear combination of sigma matrices:
\be
g_{(\theta,\phii)}^{-1}dg_{(\theta,\phii)}=
\frac{i}{2}
\Big[
(\cos\phii\sin\theta\,d\phii+\sin\phii\,d\theta)\sigma_1
 +(\sin\phii\sin\theta\,d\phii-\cos\phii\,d\theta)\sigma_2
+(1-\cos\theta)d\phii\,\sigma_3
\Big].
\nn
\ee
We can now evaluate the Berry connection (\ref{ss24}) associated with the states $\cU[g_{(\theta,\phii)}]|j\ket$ and pulled back on the sphere: 
\be
\big(A_j\big)_{(\theta,\phii)}
=
\bra j|i\cu[g_{(\theta,\phii)}^{-1}dg_{(\theta,\phii)}]|j\ket
=
\bra j|i\cu\Big[
\frac{i}{2}(1-\cos\theta)d\phii\,\sigma_3+(\cdots)\sigma_1+(\cdots)\sigma_2
\Big]|j\ket.
\label{tt38}
\ee
Here the coefficients multiplying $\sigma_1$ and $\sigma_2$ are unimportant: using eq.\ (\ref{ss33}) we can rewrite the operator $i\cu[\cdots]$ as a linear combination of $\hat S{}_j$'s; since the expectation values of $\hat S{}_1$ and $\hat S{}_2$ vanish, only the term proportional to $\sigma_3$ survives and we find the Berry connection
\be
\big(A_j\big)_{(\theta,\phii)}
=
-
\bra j|
\hat\bS{}_3
|j\ket
(1-\cos\theta)d\phii
=
-j(1-\cos\theta)d\phii
\label{s38}
\ee
which is smooth everywhere except at the South pole. Up to an overall factor this is the potential of the standard volume form on the unit sphere: the Berry curvature is
\be
F_j
=
dA_j
=
-j\sin\theta\,d\theta\wedge d\phii.
\label{ss38}
\ee
It follows that the Berry phase picked up by the states $\cU[g_{(\theta,\phii)}]|j\ket$ along a closed path on the unit sphere is the signed area of the region enclosed by the path, multiplied by $-j$:
\be
B_j[\gamma]
\refeq{s20}
\oint_{\gamma}A_j
=
\int_{\Sigma}dA_j
\refeq{ss38}
-j\times\text{area}(\Sigma),
\label{s39}
\ee
where $\Sigma$ is any oriented surface on $S^2$ such that $\der\Sigma=\gamma$.\\

Some comments are in order regarding this standard result. First, the choice of section (\ref{s37b}) is not unique since we can multiply it from the right by any rotation $h_{(\theta,\phii)}$ around the vertical axis (\ie a $(\theta,\phii)$-dependent exponential of $\sigma_3$). The corresponding Berry connection then differs from (\ref{s38}) by a $\un$ gauge transformation but the associated Berry curvature (\ref{ss38}), and hence the Berry phase (\ref{s39}), are unchanged. Our second remark concerns the orbit method. In accordance with the general results mentioned in section \ref{s27}, the parameter space $S^2=\SU/S^1$ can be interpreted as the coadjoint orbit of the coherent state $|j\ket$. Then the Berry curvature (\ref{ss38}) coincides with the orbit's symplectic form and the Berry phase (\ref{s39}) can be seen as a symplectic flux. Finally, note that for any closed path $\gamma$ on $S^2$ there exist {\it two} surfaces with boundary $\gamma$; at first sight this leads to an ambiguity in the value of (\ref{s39}) since the difference between the two results is
\be
\Delta B_j
=
j\times\text{area}(S^2)
=
4\pi j.
\label{s42}
\ee
In practice, Berry phases are only observable insofar that they appear in an exponential $e^{iB}$, so the ambiguity (\ref{s42}) is invisible as long as $j$ is an integer or a half-integer. Thus we recover standard spin quantization from the requirement that group-theoretic Berry phases be well-defined. From the point of view of geometric quantization, this is the condition that the Kirillov-Kostant symplectic form on $\SU/S^1$ be integral.

\section{Thomas precession in AdS$_3$}
\label{appa}

Here we apply the considerations of section \ref{ss15} to $\SL$. Up to topological details the latter is half of the isometry group of AdS$_3$, so its irreducible unitary representations describe particles in AdS$_3$. Since the Hamiltonian of a particle depends on its reference frame, one may investigate the Berry phases that appear when this frame changes adiabatically in a cyclic way; these phases are synonymous with Thomas precession \cite{aravind1997wigner} and provide a finite-dimensional analogue of the Virasoro Berry phases of section \ref{s63}.\\

Let us first recall some useful facts about $\SL$, the group of real $2\times2$ matrices with unit determinant (see \eg \cite[sec.\ 2.2]{Oblak:2015qia} or \cite[sec.\ 5.3.4]{Oblak:2016eij} for details). Its Lie algebra $\sl$ consists of real, traceless $2\times2$ matrices and is generated by the basis elements
\be
t_0=\demi\bmm 0 & 1 \\ -1 & 0 \emm=\frac{i}{2}\sigma_2,
\quad
t_1=\demi\bmm 0 & 1 \\ 1 & 0 \emm=\demi\sigma_1,
\quad
t_2=\demi\bmm 1 & 0 \\ 0 & -1 \emm=\demi\sigma_3.
\label{s44}
\ee
Their Lie brackets are $[t_{\mu},t_{\nu}]=\epsilon_{\mu\nu}{}^{\rho}\,t_{\rho}$ with $\epsilon_{012}=1$, all indices being raised and lowered with the three-dimensional Minkowski metric $\eta_{\mu\nu}=\text{diag}(-1,+1,+1)$. Every $\sl$ matrix can be written as a real linear combination $X=X^{\mu}t_{\mu}$. Furthermore the $\sl$ algebra admits a non-degenerate invariant bilinear form
\be
(X,Y)
\equiv
2\,\text{Tr}(XY)
=
2X^{\mu}Y^{\nu}\,\text{Tr}(t_{\mu}t_{\nu})
=
\eta_{\mu\nu}X^{\mu}Y^{\nu}.
\nn
\ee
This implies that the set of $\sl$ matrices of the form $fXf^{-1}$, where $X=X^{\mu}t_{\mu}\in\sl$ is fixed and $f$ runs over $\SL$, coincides with the orbit of the three-dimensional `energy-momentum vector' $(X^0,X^1,X^2)$ under Lorentz transformations. For instance, the adjoint orbit of $X=t_0$ is a two-dimensional hyperbolic plane
\be
\HH^2
=
\Big\{
\sqrt{1+p^2+q^2}\,t_0+p\,t_1+q\,t_2
\Big|
(p,q)\in\RR^2
\Big\},
\label{s47}
\ee
which is diffeomorphic to $\SL/S^1$ since the stabilizer of $t_0$ is the $\un$ group of rotations. This correspondence between $\SL$ and Lorentz transformations is due to the isomorphism $\SL/\ZZ_2\equiv\PSL\cong\text{SO}(2,1)^{\uparrow}$, which says that $\SL$ is the double cover of the connected Lorentz group in three dimensions. Following again the conventions of \cite{Oblak:2015qia}, this isomorphism is obtained by noting that the adjoint action of $\SL$ on its Lie algebra coincides with the vector representation of the Lorentz group in three dimensions. More precisely, for any matrix $f\in\SL$ and any $X=X^{\mu}t_{\mu}\in\sl$ (with basis elements $t_{\mu}$ given by (\ref{s44})), one can define a $3\times 3$ matrix $F[f]$ by
\be
f\,X^{\mu}t_{\mu}\,f^{-1}
=
F[f]^{\mu}{}_{\nu}X^{\nu}t_{\mu}\,.
\nn
\ee
This corresponds to a map
\be
F:\SL\rightarrow\text{SO}(2,1)^{\uparrow}:\bmm a & b \\ c & d \emm\mapsto F\left[\bmm a & b \\ c & d \emm\right]
\nn
\ee
whose explicit form turns out to be
\be
F\left[\bmm a & b \\ c & d \emm\right]
=
\bmm
\demi(a^2+b^2+c^2+d^2) & \demi(a^2-b^2+c^2-d^2) & -ab-cd\,\\[.1cm]
\demi(a^2+b^2-c^2-d^2) & \demi(a^2-b^2-c^2+d^2) & -ab+cd\,\\
-ac-bd & bd-ac & ad+bc
\emm.
\label{mama}
\ee
One can show (see \eg \cite{Oblak:2015qia}) that $F$ is a surjective homomorphism whose kernel consists of the identity matrix and its opposite, which implies that $\SL$ is indeed the double cover of $\text{SO}(2,1)^{\uparrow}$.\\

In any unitary representation $\cU$ of $\SL$, the operators $\cu[t_{\mu}]$ representing the generators (\ref{s44}) are anti-Hermitian. It is customary to define the operators
\be
L_0\equiv-i\cu[t_0],
\quad
L_1\equiv i\cu[t_1]+\cu[t_2],
\quad
L_{-1}\equiv i\cu[t_1]-\cu[t_2]
\label{s45}
\ee
which satisfy $L_m^{\dagger}=L_{-m}$ and whose commutators are $[L_m,L_n]=(m-n)L_{m+n}$ for $m,n=-1,0,1$. Then a highest weight representation of $\sl$ is built by demanding that its Hilbert space $\sH$ admit a highest weight state $|h\ket$ such that
\be
L_0|h\ket=h|h\ket,
\quad
L_1|h\ket=0
\label{hawe}
\ee
and generating the rest of $\sH$ with descendant states $(L_{-1})^n|h\ket$, where $n\geq0$.\footnote{As usual the terminology is backwards: $|h\ket$ is actually a {\it lowest}-weight state in the representation.} Assuming $\bra h|h\ket=1$, the representation is unitary if and only if $h\geq0$ (and trivial if $h=0$). The corresponding (generally projective) group representation $\cU$ is known as a discrete series representation of $\SL$.\\

Now consider one such representation $\cU$ with highest weight $h$. What one normally calls `the Hamiltonian' is the operator $L_0$, which generates time translations in a suitable reference frame. But this choice is arbitrary: one may just as well move to a different frame by a transformation $f\in\SL$ and declare that the Hamiltonian is $\cU[f]L_0\cU[f]^{-1}$. As in section \ref{ss21} this defines a family of Hamiltonians labelled by $f$, and one can study the resulting Berry phases. Let us focus on the highest weight state $|h\ket$; it is an eigenvector of the `standard' Hamiltonian $L_0$ and its stabilizer is the $\un$ group of rotations, but generic $\SL$ transformations act on it non-trivially: the set of inequivalent states that can be reached from $|h\ket$ is a manifold $\HH^2\cong\SL/S^1$. In terms of the global coordinates $(p,q)$ in (\ref{s47}), any transformation $f$ that maps the tip of the hyperboloid on a point $(p,q)$ also maps $|h\ket$ on a new state $\cU[f]|h\ket$ which has definite energy $h$ with respect to the `non-standard' Hamiltonian $\cU[f]L_0\cU[f]^{-1}$. Our goal is to find the Berry phase picked up by these states as they undergo a family of boosts that trace a closed path on the hyperbolic plane (\ref{s47}). (Such phases were studied \eg in \cite{berry1985classical} and their classical version has been observed in several experiments \cite{kitano1989observation}.)\\

As in (\ref{s36}), to describe the states $|h;p,q\ket$ we need to choose a family of transformations $g_{(p,q)}\in\SL$, each mapping the tip of (\ref{s47}) on the point $(p,q)$, such that
\be
|h;p,q\ket
=
\cU[g_{(p,q)}]|h\ket.
\nn
\ee
One can think of this family as a section of the $\un$ bundle $\SL\rightarrow\HH^2$, associating a group element $g_{(p,q)}$ with a point $(p,q)$. In contrast to the $\SU$ case, the bundle is now trivial and global smooth sections exist. Writing $E\equiv\sqrt{1+p^2+q^2}$, we take
\be
g_{(p,q)}
=
\frac{1}{\sqrt{2}}
\bmm
\big[E+1\big]^{1/2}+\frac{p}{\sqrt{p^2+q^2}}\big[E-1\big]^{1/2}
&
-\frac{q}{\sqrt{p^2+q^2}}\big[E-1\big]^{1/2}\\[.5cm]
-\frac{q}{\sqrt{p^2+q^2}}\big[E-1\big]^{1/2}
&
\big[E+1\big]^{1/2}-\frac{p}{\sqrt{p^2+q^2}}\big[E-1\big]^{1/2}
\emm.
\label{ss48}
\ee
Here the normalization ensures that the adjoint action of $g_{(p,q)}$ maps the tip of (\ref{s47}) on the point with coordinates $(p,q)$; indeed, applying (\ref{mama}) to the $\SL$ group element (\ref{ss48}), one gets
\be
F[g_{(p,q)}]
=
\bmm
\sqrt{1+p^2+q^2} & \times & \times \\
p & \times & \times \\
q & \times & \times
\emm
\label{mamas}
\ee
where the crossed entries are irrelevant for our purposes: when acting on the tip $(1,0,0)^t$ of (\ref{s47}), the boost (\ref{mamas}) maps it on the point $(\sqrt{1+p^2+q^2},p,q)^t$ as desired.\\

The section (\ref{ss48}) can be used to pullback the Maurer-Cartan form (\ref{t24}) from $\SL$ to $\HH^2$. The result is an $\SL$ gauge field that can be expressed as a linear combination
\be
g_{(p,q)}^{-1}dg_{(p,q)}=X^{\mu}t_{\mu}
\nn
\ee
with components $X^{\mu}$ given by
\be
X^0
=
\frac{E-1}{p^2+q^2}(p\,dq-q\,dp),
\quad
X^1+iX^2
=
idp-dq-\frac{E-1}{E}\frac{ip-q}{p^2+q^2}(p\,dp+q\,dq).
\nn
\ee
We can now evaluate the Berry connection (\ref{ss24}) associated with the states $\cU[g_{(p,q)}]|h\ket$ and pulled back on the hyperbolic plane:
\be
\big(A_h\big)_{(p,q)}
=
\bra h|i\cu[g_{(p,q)}^{-1}dg_{(p,q)}]|h\ket
=
\bra h|i\cu\bigg[
\frac{E-1}{p^2+q^2}(pdq-qdp)t_0+(\cdots)t_1+(\cdots)t_2
\bigg]|h\ket.
\nn
\ee
This is the $\SL$ analogue of eq.\ (\ref{tt38}); as in the latter case the coefficients of $t_1$ and $t_2$ are unimportant since they multiply linear combinations of the operators $L_{\pm 1}$ whose expectation values vanish by virtue of (\ref{hawe}). Hence only the term proportional to $t_0$ survives and we get the Berry connection
\be
A_h
=
-\bra h|
L_0
|h\ket
\frac{E-1}{p^2+q^2}(p\,dq-q\,dp)
\refeq{hawe}
-h\frac{\sqrt{1+p^2+q^2}-1}{p^2+q^2}(p\,dq-q\,dp)
\nn
\ee
where we used (\ref{s45}) and rewrote $E$ in terms of $p,q$. Up to normalization this is the potential of the standard Lorentz-invariant volume form on $\HH^2$: the Berry curvature is
\be
F_h
=
dA_h
=
-h\,\frac{dp\wedge dq}{\sqrt{1+p^2+q^2}}.
\label{ss49}
\ee
It follows that the Berry phase picked up by the states $\cU[g_{(p,q)}]|h\ket$ along a closed path $\gamma$ in $\HH^2$ is the signed area of the enclosed region, multiplied by $-h$:
\be
B_h[\gamma]
=
\oint_{\gamma}A_h
=
\int_{\Sigma}dA_h
=
-h\times\text{area}(\Sigma),
\label{s50}
\ee
where $\Sigma$ is the oriented surface with finite area on $\HH^2$ such that $\der\Sigma=\gamma$. All these results are directly analogous to those of the $\SU$ case, and essentially the same comments apply. In particular one can verify that the curvature (\ref{ss49}), and hence the phase (\ref{s50}), are independent of the choice of section (\ref{ss48}). Furthermore these quantities have a natural interpretation in the orbit method: the parameter space $\HH^2=\SL/S^1$ is the coadjoint orbit of the coherent state $|h\ket$ and the Berry curvature coincides with the Kirillov-Kostant symplectic form. Note that in contrast to $\SU$ there is no ambiguity of the type (\ref{s42}) for $\SL$, hence no quantization condition on the weight $h$. (The weight does get quantized if one insists that the representation of $\SL$ be non-projective, but this is due to topological, rather than symplectic, issues.)\\

For comparison with the Virasoro case, let us evaluate the Berry phases (\ref{s50}) associated with circles around the tip of $\SL/S^1$. They are most easily described with polar coordinates $(r,\phii)$ such that $p=r\cos\phii$, $q=r\sin\phii$, in terms of which the Berry curvature (\ref{ss49}) reads
\be
F_h
=
-h\,\frac{r\,dr\wedge d\phii}{\sqrt{1+r^2}}.
\nn
\ee
A circle around the tip is a curve where $r(t)=R$ is constant while $\phii(t)=\omega t$ for some angular velocity $\omega>0$. The time needed to perform one turn around the tip is $T=2\pi/\omega$ and the corresponding Berry phase (\ref{s50}) is
\be
B_h(R)
=
-2\pi h\big(\sqrt{1+R^2}-1\big)
=
-2\pi h(\text{cosh}\lambda-1)
\equiv
B_h(\lambda),
\nn
\ee
where in the second equality we have introduced the rapidity $\lambda\equiv\text{argsinh}(R)$. This reproduces the formula (\ref{s53}) displayed above.

\addcontentsline{toc}{section}{References}

\end{document}